%% file: Comp_NonL.tex
\documentclass[conference]{IEEEtran}
\usepackage{tikzfig}
\input{tikzfigures.tex}
\input{tikzstyles.tex}

\usepackage{times}
\usepackage{epsfig}
\usepackage{amssymb}
\usepackage{amsmath}
\usepackage{latexsym}
\usepackage{epsfig}
\usepackage{color}
\usepackage{verbatim}
\usepackage{graphicx}
\usepackage{xspace}

\input{defs.tex}

\def\bR{\begin{color}{red}} 
\def\bB{\begin{color}{blue}}
\def\bM{\begin{color}{magenta}}
\def\bC{\begin{color}{cyan}}
\def\bW{\begin{color}{white}}
\def\bBl{\begin{color}{black}} 
\def\bG{\begin{color}{green}}
\def\bY{\begin{color}{yellow}}
\def\e{\end{color}}

\newtheorem{Th}{Theorem}[section] 
\newtheorem{theorem}[Th]{Theorem} 
\newtheorem{proposition}[Th]{Proposition} 
\newtheorem{lemma}[Th]{Lemma} 
\newtheorem{corollary}[Th]{Corollary} 
\newtheorem{definition}[Th]{Definition} 
\newtheorem{example}[Th]{Example}

\begin{document}
\title{Strong Complementarity and Non-locality \\in Categorical Quantum Mechanics}
\author{
\IEEEauthorblockN{Bob~Coecke${}^1$ \quad Ross Duncan${}^2$ \quad Aleks Kissinger${}^1$ \quad Quanlong Wang${}^3$}
\IEEEauthorblockA{\\
${}^1$University of Oxford, Department of computer science,\\ 
Wolfson Building, Parks Road, Oxford OX1 3QD, UK. \\ 
{\em coecke/alek@cs.ox.ac.uk}\\
${}^2$Universit\'e Libre de Bruxelles, Laboratoire d'Information
Quantique\\
Campus Plaine, Boulevard du Triomphe, 1050 Brussels, Belgium\\
{\em rduncan@ulb.ac.be}\\
${}^3$Beihang University, School of Mathematics and System Sciences\\
XueYuan Road No.37, HaiDian District, Beijing, China\\ 
{\em qlwang@buaa.edu.cn}
%\\ (authors in alphabetical order)
} %
}

\maketitle

%%%%% TODO: why do we use complementary measurements, are these all of them?

\begin{abstract}
Categorical quantum mechanics studies quantum theory in the framework of dagger-compact closed categories.

Using this framework, we establish a tight relationship between two
key quantum theoretical notions: non-locality  and complementarity.
In particular, we establish a direct connection between Mermin-type
non-locality scenarios, which we  generalise to an arbitrary number of
parties, using systems of arbitrary dimension, and performing
arbitrary measurements,  and,  a new stronger notion of
complementarity which we introduce here. 

Our derivation of the fact that strong complementarity is a necessary
condition for a Mermin scenario provides a crisp operational
interpretation for strong complementarity. We also provide a complete classification of strongly complementary observables for quantum theory, something which has not yet been achieved for ordinary complementarity. 

Since our main results are expressed in the (diagrammatic) language of
dagger-compact categories, they can be applied outside of quantum
theory, in any setting which supports the purely algebraic notion
of strongly complementary observables.  We have therefore introduced a
method for discussing non-locality in a wide variety of models in
addition to quantum theory. 

The diagrammatic calculus substantially simplifies (and sometimes even trivialises) many of the derivations, and provides new insights. In particular, the diagrammatic computation of correlations clearly shows how local measurements interact to yield a global overall effect.  In other words, we \em depict non-locality\em.
\end{abstract}

\section{Introduction}

This paper is concerned with two central notions in quantum foundations and quantum computation,
 \em non-locality \em and \em complementarity\em, and the study
of their relationship.

Non-locality is what Einstein notoriously referred to as `spooky
action at a distance'. It was formally substantiated for the first
time by Bell's theorem, and experimentally verified by testing
Bell-inequalities.  It states that the correlations observed when
measuring spatially quantum separated systems cannot be explained by
means of classical probabilities, i.e.~that there is no underlying \em
hidden variable theory\em.  Complementarity, informally put, states
that if one knows the value of one observable sharply (e.g.~position),
then there is complete uncertainty about the other observable
(e.g.~momentum).

These two concepts underpin what is arguably the most successful
endeavour towards quantum information technologies: quantum
cryptography. Indeed, in a quantum key distribution protocol,
encoding data in either of two complementary observables will enable
the parties to detect interception by an adversary \cite{BB84}, while
non-locality allows one to verify the authenticity of the shared
entangled resource by means of which key sharing is established
\cite{Ekert91}.

%This paper contributes in two distinct manners to the area of computer science logic:
%\begin{itemize}
%\item It provides categorical semantics for quantum computation, hence exposing logical aspects thereof.
%\item It relies on CS-methods to obtain new results in fundamental physics, and applications thereof.
%\end{itemize}
This research applies methods of computer science and logic to investigations in quantum foundations.
It is moreover  strongly aligned with the current trend in the broader quantum information community: understanding quantum information processing within a larger space of hypothetical information processing theories in order to understand what is particular about quantum theory.

Until now, this area has been characterised by the study of \em generalised probabilistic theories
\em \cite{NLB}, a space of theories which includes quantum
probability theory, classical probability theory, as well as theories
which are even more non-local than quantum theory.  A topic of
particular focus has been the search for peculiarities of quantum
non-locality, within the larger space of non-local theories,
e.g.~\cite{InfoCaus}, \cite{InfoCaus2}. 

Of course, a space of more general theories can be conceived as
abstracting away certain concrete features of quantum theory, hence
encompassing a broader class of theories, and the words `generalised'
and `abstract' can be treated as synonymous.  While the generalised
probabilistic theories discussed above abstract away all but convex
probabilistic structure, our focus is on the \em compositional
structure on processes\em, say \em generalised process theories\em.

While having composition play a leading role evidently draws form
computer science practice, it also appeals to Schr\"odinger's
conviction that what mostly characterises quantum theory is the manner
in which systems compose \cite{Schrodinger}.

This compositional paradigm was the main motivation for  \em categorical quantum mechanics \em (CQM), initiated by Abramsky and Coecke in \cite{AC}. Meanwhile, CQM has helped to solve open problems in Quantum Information and Computation e.g.~\cite{CK,DP,Horsman}, and has meanwhile been adopted by leading researchers  in the area of quantum foundations e.g.~\cite{Chiri}, \cite{Hardy}.

In this paper we push CQM far beyond its previously established
horizons, both on the level of comprehensiveness and in terms of its
application domain. The vehicle to do so is Mermin's ingenious but
elaborate non-locality proof \cite{Harvey, Harveybis, GHZ, Mermin}.
This argument, usually stated mostly in natural language,  establishes
non-locality as a \em contradiction of parities \em for quantum theory
and local theories, rather than as the violation of a Bell-inequality,
and involves a sophisticated interplay of several incompatible
measurement scenarios, each involving measurements against varying
angles, as well as hypothetical hidden variable theories, and the
manipulations of the resulting probabilistic measurement data.

Our abstraction as well as our generalisations of this scenario
provides important new physical insights in the nature of
non-locality, and its relationship to complementarity, as we discuss
in Section \ref{sec:outlook}. Hence this paper is both one in Computer
Science as well as one in Quantum Foundations.

\medskip
{\bf Outline.} Sec.~\ref{Sec:BackgrProcTheor}  provides an overview of CQM and presents  some relevant non-standard models.  We  briefly recall the diagrammatic calculus for symmetric monoidal categories.

In Sec.~\ref{Sec:complementarity} we define strong complementarity, establish its relationship with ordinary complementarity, and state the first major result of this paper: the classification of strongly complementary observables in quantum theory.

In Sec.~\ref{sec:corcomp} we diagrammatically compute the correlations of measurements against arbitrary angles on an $n$-party GHZ-state for systems of arbitrary type.

In Sec.~\ref{sec:GHZMerminAbs} we cast the usual Mermin argument within CQM, as a stepping stone to its generalisation in Sec.~\ref{sec:Mermin_general}. In doing so we rely crucially on strong complementarity.

Finally, in Sec.~\ref{sec:converse} we establish the necessity of strong complementarity for Mermin arguments. This also provides an operational interpretation for strong complementarity. 

\medskip
{\bf Earlier work.} In \cite{CD,CD2} Coecke and Duncan introduced the graphical \em \textsc{zx}-calculus \em  for the specific purpose of reasoning graphically about qubits. This calculus included the equations that we identify here as strong complementarity, but these were not identified as such.  In \cite{CES} Coecke, Edwards and Spekkens relied on  the CQM concept of a phase groups to identify the differences of the categories ${\bf Stab}$ and ${\bf Spek}$ with respect to non-locality, but the Mermin scenario was not formulated within CQM. In \cite{Bill, Bill2} Edwards expanded this line work and derived some higher-dimensional generalisations of the Mermin argument.

\section{Background I }\label{Sec:BackgrProcTheor} 

\subsection{Models of Physical Theories}

Symmetric monoidal categories (SMCs) provide a very general language
for physical theories: a morphism $f:A\to B$ is interpreted as a
process from a physical system of type $A$ to one of type $B$.
Spatial and temporal extension are represented by the composition and
tensor product; that is, by the sequential and parallel combination of
processes.  A \em state \em of system $A$ is simply a morphism
$\psi:I\to A$, while an \em effect \em has type $\pi:A\to I$.  Every
SMC has a commutative monoid of \em scalars\em, the
morphims  $s:I\to I$.

Each concrete SMC is a model of this primitive theory, with different
physical characteristics.  For example, $({\bf FHilb}, \otimes)$, the
category of of finite dimensional Hilbert spaces and linear maps,
comprises quantum systems and pure post-selected quantum
processes\footnote{`Pure' as in `not mixed'; `post-selected' means
  that processes which are conditional on obtaining a given
  measurement outcome are permitted.}.  We write ${\bf FHilb}_D$ to
denote the category of Hilbert spaces of dimension $D^n$, for some
fixed $D$.  In particular, ${\bf Qubit}:={\bf FHilb}_2$.  In fact,
${\bf FHilb}$ has redundencies:
two linear maps that are equal up to a non-zero scalar multiple
represent the same physical process.

This redundancy can be eliminated by using a different model.  The
category $CP({\bf FHilb})$ models open quantum systems and
completely positive maps.  It is built from ${\bf FHilb}$ using
Selinger's $CP$-construction \cite{SelingerCPM}, described
Sec.~\ref{sec:clas-quant}, which constructs a
category of \emph{mixed} processes from any given category of pure
processes.

Despite its seemingly different nature, $({\bf FRel}, \times)$, the
category of finite sets and relations with the Cartesian product as
the tensor is a key example.  Restricting to
powers of a fixed set of size $D$ we write ${\bf FRel}_D$.

In fact, ${\bf FRel}$ shares many features of ${\bf FHilb}$ and
$CP({\bf FHilb})$.  All three  are  \em
compact \em categories \cite{KellyLaplaza}:  each object $A$ has
a dual object $A^*$ and morphisms $\eta_A: I\to A^*\otimes A$ and
$\varepsilon_A: A\otimes A^*\to I$ such that  $(\varepsilon_A \otimes
1_A)\circ (1_A\otimes \eta_A)=1_A$ and $(1_{A^*} \otimes
\varepsilon_A)\circ(\eta_A\otimes 1_{A^*})=1_{A^*}$.  We refer to
these elements as the \emph{compact structure} on $A$.  Note that they need
not be unique: an object may support several compact structures.

Further, our three examples are all \em dagger compact \em
\cite{AC}: there exists an identity-on-objects involutive
contravariant strict monoidal endofunctor $\dagger: {\bf C}\to {\bf
  C}$ with $\varepsilon_A=\eta_A^\dagger\circ \sigma_{A^*, A}$ for all
objects $A$, to which we refer as the \em dagger functor\em.

Dagger compactness provides the language for many quantum concepts
such as Bell-states/effects, unitarity, the Born rule \cite{AC}, and complete
positivity \cite{SelingerCPM}.  E.g.~\em unitarity \em means
$U^\dagger\circ U=1_A$ and $U\circ U^\dagger=1_B$. An SMC with a
dagger functor is referred to as $\dagger$-SMC. In any $\dagger$-SMC
we can define an inner product for states $\psi$ and $\phi$ as
$\psi^\dagger\circ\phi:I\to I$.  This provides the usual inner-product
(in ${\bf FHilb}$) as well as amplitudes (in $CP({\bf FHilb})$). More
examples of physical models can be found in Appendix~\ref{sec:more-models}.

\subsection{$\dagger$-Compact Categories and Diagrammatic Calculus}\label{sec:diagram}

Monoidal categories admit a diagrammatic notation which greatly
simplifies the task of reading, analysing, and computing in this
framework.  For the $\dagger$-compact categories of interest here this
language takes a particularly simple form. 

Systems (objects) are depicted by labelled wires, and processes
(morphisms) are represented as boxes with wires in and out, indicating
the type of the process. Composition is expressed by plugging the
outputs of one box into the inputs of another, and the monoidal product
is given by juxtaposition. The monoidal unit $I$ is represented as the
empty diagram. 
\ctikzfig{cat_compose_tensor}

We shall draw composition from bottom-to-top, and omit wire labels
where there is no ambiguity. In a symmetric monoidal category, we
indicate the swap map as a wire crossing. For compact closed
categories, we indicate a dual object $A^*$ by a wire of the opposite
direction. The cap and cup maps are then depicted as half-turns. 
\ctikzfig{cat_cap_cup}

Using caps and cups, we can turn any morphism $f : A \rightarrow B$
into a morphism on the dual objects going in the opposite direction:
$f^* : B^* \rightarrow A^*$. 
\ctikzfig{bend_wires}

This is sometimes called the transpose of $f$, but this terminology
can be misleading. In $\catFHilb$, $L^*$ is the map that takes a
linear form $\bra\xi \in B^*$ to $\bra\xi\!L \in A^*$. We refer to
this map simply as the \textit{upper-star} of $f$. In a
$\dagger$-category, the $\dagger$ functor sends $f : A \rightarrow B$
to $f^\dagger : B \rightarrow A$. In a $\dagger$-compact category, we
define the \textit{lower-star} of $f$ as $f_* := (f^\dagger)^* =
(f^*)^\dagger$. 

Since our category  is symmetric, wires are allowed to cross, and boxes
may slide up along wires freely without changing the denotation of the
diagram.  More generally, if one diagram may be deformed continuously
to another, then these diagrams denote the same arrow in the category.
For a more comprehensive description of graphical languages, see
Selinger~\cite{SelingerSurvey}. 

In the diagrams to come, we will often use horizontal separation to
indicate separation in space and vertical separation to indicate
separation in time. For example,  
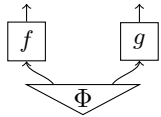
\[
\beginpgfgraphicnamed{spacetime}
\InputIfFileExists{spacetime.tikz}{}{\input{./figures/spacetime.tikz}}
\endpgfgraphicnamed 
\]
depicts the creation of two systems by the process $\Phi$, which then
become spatially separated over some time and are acted upon by
processes $f$ and $g$ respectively.  

\subsection{Generalised Observables}\label{sec:observables}

An \emph{observable} yields classical data from a physical system
\cite{CPav, CPaqPav}. In quantum mechanics, an observable is
represented by a self-adjoint operator. The important information
encoded by a (non-degenerate) observable is its orthonormal basis of
eigenstates. In ${\bf FHilb}$, orthonormal bases (ONBs) are in 1-to-1
correspondence with $\dagger$-special commutative Frobenius
algebras~\cite{CPV}. 

\begin{definition}
  In a $\dagger$-SMC, a $\dagger$-special Frobenius algebra ($\dagger$-SCFA) is a commutative Frobenius algebra
  \begin{align*}
  \mathcal O_{\!\smallwhitedot} = ( &
    \whitemu : X \otimes X \rightarrow X, \ \ 
    \whiteeta : I \rightarrow X, \\
  & \whitedelta : X \rightarrow X \otimes X, \ \ 
    \whiteepsilon : X \rightarrow I)
  \end{align*}
  such that $\whitedelta = (\whitemu)^\dagger$, $\whiteepsilon = (\whiteeta)^\dagger$ and
\beginpgfgraphicnamed{special}
\InputIfFileExists{special.tikz}{}{\input{./figures/special.tikz}}
\endpgfgraphicnamed.
\end{definition}

Because the correspondence to ONBs, a $\dagger$-SCFA is also called an
\textit{observable structure}. We will use the symbolic representation
$(\whitemu, \whiteeta, \whitedelta, \whiteepsilon)$ and the pictorial
one $(\whitemult, \whiteunit, \whitecomult, \whitecounit)$
interchangeably. 

The key distinction between classical and quantum data is that
classical data may be freely copied and deleted while this is
impossible for quantum data, due to the no-cloning \cite{Dieks, WZ}
and no-deleting \cite{Pati} theorems. This is a general fact: any
compact closed category with natural diagonal maps $\Delta : X
\rightarrow X \otimes X$ collapses to a
preorder~\cite{Abramsky-no-cloning}.

\begin{proposition}\label{prop:spider}\em 
Given an observable structure $\whiteobs$ on $X$, let
$(\whitedot)_n^m$ denote the `$(n,m)$-legged spider':
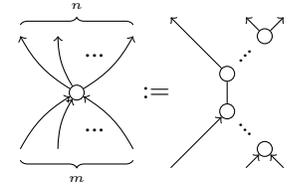
\[
 \begin{tikzpicture}
    \begin{pgfonlayer}{nodelayer}
        \node [style=none] (0) at (-2, 1) {};
        \node [style=none] (1) at (-0.5, 1) {};
        \node [style=none] (2) at (0, 1) {};
        \node [style=none] (3) at (1, 1) {};
        \node [style=none] (4) at (1.5, 1) {};
        \node [style=none] (5) at (-2, 0.75) {};
        \node [style=none] (6) at (-1.5, 0.75) {};
        \node [style=none] (7) at (-0.5, 0.75) {};
        \node [style=white dot] (8) at (1.25, 0.75) {};
        \node [style=none] (9) at (-1, 0.5) {...};
        \node [style=whitebg] (10) at (1, 0.5) {\small \rotatebox[origin=c]{45}{...}};
        \node [style=white dot] (11) at (0.75, 0.25) {};
        \node [style=white dot] (12) at (-1.25, 0) {};
        \node [style=none, anchor=east] (13) at (0, 0) {$:=$};
        \node [style=white dot] (14) at (0.75, -0.25) {};
        \node [style=none] (15) at (-1, -0.5) {...};
        \node [style=whitebg, fill=white] (16) at (1, -0.5) {\small \rotatebox[origin=c]{-45}{...}};
        \node [style=none] (17) at (-2, -0.75) {};
        \node [style=none] (18) at (-1.5, -0.75) {};
        \node [style=none] (19) at (-0.5, -0.75) {};
        \node [style=white dot] (20) at (1.25, -0.75) {};
        \node [style=none] (21) at (-2, -1) {};
        \node [style=none] (22) at (-0.5, -1) {};
        \node [style=none] (23) at (0, -1) {};
        \node [style=none] (24) at (1, -1) {};
        \node [style=none] (25) at (1.5, -1) {};
    \end{pgfonlayer}
    \begin{pgfonlayer}{edgelayer}
        \draw [style=diredge, bend left=15] (18.center) to (12);
        \draw (14) to (11);
        \draw [style=diredge] (11) to (2.center);
        \draw [style=small braceedge] (22.center) to node[wire label, inner sep=3 pt]{$m$} (21.center);
        \draw [style=diredge] (8) to (4.center);
        \draw [style=diredge] (23.center) to (14);
        \draw [style=diredge] (24.center) to (20);
        \draw [style=diredge] (8) to (3.center);
        \draw [style=diredge, bend left=15] (12) to (5.center);
        \draw [style=diredge, bend right=15] (19.center) to (12);
        \draw [style=diredge, bend left=15] (17.center) to (12);
        \draw [style=diredge, bend right=15] (12) to (7.center);
        \draw [style=diredge] (25.center) to (20);
        \draw [style=diredge, bend left=15] (12) to (6.center);
        \draw [style=small braceedge] (0.center) to node[wire label, inner sep=3 pt]{$n$} (1.center);
    \end{pgfonlayer}
\end{tikzpicture} \,;
\]
then any morphism $X^{\otimes n}\to X^{\otimes m}$ built from
$\whitemu, \whiteeta, \whitedelta$ and $\whiteepsilon$ via $\dagger$-SMC structure which has a
connected graph is equal to the $(\whitedot)^m_n$. Hence, spiders
compose as follows:
\begin{equation}\label{eq:spidercomp}
\beginpgfgraphicnamed{spidercomp}
\InputIfFileExists{spidercomp.tikz}{}{\input{./figures/spidercomp.tikz}}
\endpgfgraphicnamed
\end{equation}
\end{proposition}
 
% \begin{remark}
% Non-degenerate observables can also be defined as in \cite{CPav, CPaqPav}, but won't play a role in this paper.
% % \bR ... general observables ones arise as $\dagger$-modules, or equivalently, 
% Formally, they are $\dagger$-Eilenberg-Moore coalgebras for the  comonad $X\otimes-$ induced by the comonoids of Defn.~\ref{def:observable}.
% \end{remark} 
 
Concretely, given an ONB $\{|i\rangle\}_i$ then
$\whitedelta:: |i\rangle\mapsto |ii\rangle$ defines an observable, and
all observables are of this form for some ONB.  The resulting
intuition is that $\whitedelta$ is an operation that  `copies' basis
vectors, and  that $\whiteepsilon$  `erases' them \cite{CPaqPav}. 

% In the category $n{\bf Cob}$ there are no non-trivial  observables and in  ${\bf FSet}$ there is exactly one on each object.

%In contrast, 
Perhaps surprisingly, ${\bf FRel}$ also has many distinct observables,
which have been classified by Pavlovic \cite{Dusko}.  Even on the two
element set there are two \cite{Spek}, namely $\whitedelta: \{0,
1\}\to \{0, 1\}\times \{0, 1\}:: i\mapsto (i, i)$ and $\graydelta::
0\mapsto \{(0,0), (1,1)\}; 1\mapsto \{(0,1), (1,0)\}$.  In fact, this
pair is strongly complementary in the sense of
Sec.~\ref{Sec:complementarity}.
% Also in ${\bf Spek}$
% Defn.~\ref{def:observable} provides the intended notion of observable.

Each observable structure comes with a set of \textit{classical
  points}, the abstract analogues to eigenvectors of an observables. A
classical point is a state that is copied by its comultiplication and
deleted by the counit: 
\begin{equation}\label{eq:classpoint}
\beginpgfgraphicnamed{classpoint}
\InputIfFileExists{classpoint.tikz}{}{\input{./figures/classpoint.tikz}}
\endpgfgraphicnamed 
\end{equation}

We will depict classical points as triangles of the same colour as
their observable structure. Each observable structure furthermore
defines a \textit{self-dual} $\dagger$-compact structure on its
object. That is, it defines a $\dagger$-compact structure where the
dual object of $X$ is $X$ itself. 
By (\ref{eq:spidercomp}) we have:
\ctikzfig{frob_comp}

The upper-star with respect to this compact structure corresponds in
\catFHilb to transposition in the given basis. For that reason, we
call this the $\whitedot$-transpose $f^\whitetranspose$. The lower
star corresponds to complex conjugation in the basis of $\whiteobs$,
so we call it the $\whitedot$-conjugate $f_\whiteconjugate :=
(f^\whitetranspose)^\dagger$. 

%%%%% ALEKS : do we have to? I think "proportional to a classical point" is fine in all cases
% We will occasionally refer to points \textit{proportional} to classical points as classical points i.e.~when (\ref{eq:classpoint}) hold up to a scalar. 

%%%%% ALEKS : I don't think this is necessary, since we are careful with directions.
% Each observable on $X$  induces a $\dagger$-compact structure on $X$ with $\whiteeta:=\whitecup$ and $\whiteepsilon:= \whitecap$
% since by (\ref{eq:spidercomp}) we have:
% \[
%  \tikzfig{frob_comp} 
% \]

% This induced dagger compact structure allows one to define  the \em transpose \em and \em conjugate \em of a morphism $f: X\to Y$ relative to observables $\whitedelta$ on $X$ and  $\graydelta$ on $Y$ as: 
% \[
% f^T=\tikzfig{transpose}  : Y\to X \qquad \overline{f}=\tikzfig{conjugate}  : X\to Y\,.
% \]

%% \bR ... circle + independence of observable (no proof)  ... \e  

\subsection{Phase Group for an Observable Structure}\label{sec:phasegroup}
 
Given an observable structure $\whiteobs$ on $X$, the multiplication $\whitemu$
puts a monoid structure on the points of $X$. If we restrict to those
points $\psi_\alpha : I \to A$ where $\whitemu (\psi_\alpha \otimes
(\psi_\alpha)_\whiteconjugate) = \whiteeta$, we obtain an Abelian
group $\whitePhi$ called the \emph{phase group} of
$\whiteobs$~\cite{CD}. We let $\psi_{-\alpha} :=
(\psi_\alpha)_\whiteconjugate$ and represent these points as circles
with one output, labelled by a phase. 
\ctikzfig{phases} 

For an observable on the qubit, the phase group consists of the points
that are unbiased to the eigenstates of the observable, and their
multiplication is their convolution.  Explicitly, given $\whitemu=
\ketbra{0}{00} + \ketbra{1}{11}$, we have: 
\[
\whitemu
\left(\left(\begin{array}{c}
1\\ e^{i\alpha}
\end{array}\right) 
\otimes
\left(\begin{array}{c}
1\\ e^{i\beta}
\end{array}\right)\right)
=
\left(\begin{array}{c}
1\\ e^{i(\alpha+\beta)}
\end{array}\right)\,,
\] 
so we obtain the circle group.  
%Therefore we will denote these phases by greek letters and their multiplication by $+$. 

We can now introduce `spiders decorated with phases':
\begin{equation}\label{eq:decspider}
\beginpgfgraphicnamed{decspider}
\InputIfFileExists{decspider.tikz}{}{\input{./figures/decspider.tikz}}
\endpgfgraphicnamed
\end{equation}
which compose as follows:
\begin{equation}\label{eq:decspidercomp} 
\beginpgfgraphicnamed{decspidercomp}
\InputIfFileExists{decspidercomp.tikz}{}{\input{./figures/decspidercomp.tikz}}
\endpgfgraphicnamed
\end{equation}

The name `phase group' comes from fact that phased spiders with one
input and one output are the abstract analogue to phase gates,
familiar from quantum computing. This is supported by the following
fact, proven in~\cite{CD2}.  
\begin{proposition}\label{lem:phases-are-unitary}
  If $\psi_\alpha \in \whitePhi$ then $\whitemu \circ (1_X \otimes
  \psi_\alpha) : X\to X$ (i.e. a spider with one input and one output,
  labelled by $\alpha$) is unitary. 
\end{proposition}

%%%%% we only have one colour so far %%%%%%%%%%%%%%%%%%%
%
% Equations (\ref{eq:decspider},\ref{eq:classpoint}) and
% (\ref{eq:decspidercomp}) can be summarized in the following slogan,
% justified in Thm.~\ref{thm:subpahsegroup}: \em ``decorated points are
% `fused' by their own colour and `copied' by the other colour''\em. 

\subsection{Generalised Classical-Quantum Interaction}\label{sec:clas-quant}

In ordinary quantum theory, quantum states are represented as positive
operators and operations as completely positive maps, or CPMs. These
are maps that take positive operators to positive operators. A general
CPM can be written in terms of a set of linear maps $\{ B_i : \mathcal
H_1 \rightarrow \mathcal H_2 \}$ called its \textit{Kraus maps}. 
\[ \Theta(\rho) = \sum_i B_i \rho B_i^\dagger \]

Since \catFHilb is compact, we can regard $\rho : \mathcal H_1
\rightarrow \mathcal H_1$ as an element of $\mathcal H_1^* \otimes
\mathcal H_1$. 
\ctikzfig{rho_point} 

Then, we can encode the Kraus vectors of $\Theta$ in a map $B' = \sum
\ket{i} \otimes B_i$ and represent $\Theta$ as: 
\begin{equation}
\beginpgfgraphicnamed{kraus_decomp}
\InputIfFileExists{kraus_decomp.tikz}{}{\input{./figures/kraus_decomp.tikz}}
\endpgfgraphicnamed
\end{equation}

When we take the elements in Eq. (\ref{eq:cp-map}) to be morphisms in
an arbitrary $\dagger$-compact category, this gives us an abstract
definition of a completely positive map. This is Selinger's
representation of CPMs~\cite{SelingerCPM}. 
\begin{equation}\label{eq:cp-map}
\beginpgfgraphicnamed{abstract_cp}
\InputIfFileExists{abstract_cp.tikz}{}{\input{./figures/abstract_cp.tikz}}
\endpgfgraphicnamed
\end{equation}

Important special cases are \textit{states} where $A \cong I$,
\textit{effects} where $B \cong I$, and `pure' maps, where $C \cong
I$. 

Returning to quantum mechanics, we can see how a quantum measurement
would look in this language. A (projective) quantum measurement
$M_{\smallwhitedot}$ is a CPM that sends trace 1 positive operators
(in this case quantum states) to trace 1 positive operators that are
diagonal in some ONB (encoding a probability distribution of
outcomes). Suppose we wish to measure with respect to $\grayobs$,
whose classical points form an ONB $\{ \ket{x_i} \}$. The probability
of getting the $i$-th measurement outcome is computed using the Born
rule. 
\[ \textrm{Prob}(i, \rho) = \textrm{Tr}(\ketbra{x_i}{x_i} \rho) = \bra{x_i} \rho \ket{x_i} \]

We can encode this map from states to distributions as:
\[ M_{\smallgraydot}(\rho) = \sum_i \left(\bra{x_i} \rho \ket{x_i} \right) \ketbra{x_i}{x_i} \]

Or, rather than encoding the distribution in a diagonal matrix, we
could simply use a vector: 
\[ m_{\smallgraydot}(\rho) = \sum_i \left(\bra{x_i} \rho \ket{x_i}\right) \ket{x_i} \]

Expanding this graphically, we have:
\ctikzfig{measurement_derivation}

We are now ready to make definitions for abstract measurements and
abstract probability distributions, which we shall call Born vector. 

\begin{definition}
  For an observable structure $\grayobs$, a measurement is defined as
  the following map: 
  \ctikzfig{measurement}
  Any point $\roundket{\Gamma} : I \rightarrow X$ of the following
  form is called a \textit{Born vector}, with respect to $\grayobs$: 
  \ctikzfig{born_vector}
\end{definition}

We can naturally extend the definition above to points of the form
$\roundket{\Gamma} : I \rightarrow X \otimes \ldots \otimes X$ by
requiring that they be Born vectors with respect to the product
Frobenius algebra $\grayobs \otimes \ldots \otimes \grayobs$. 

The adjoint of the measurement map $m_{\smallgraydot}^\dagger$ is a
preparation operation. In \catFHilb, it takes a Born vector
$\roundket{\Gamma}$ with respect to $\grayobs$ and produces a
probabilistic mixture of the (pure) outcome states of $\grayobs$ with
probabilities given by $\roundket{\Gamma}$.

This leads to a simple classical vs.~quantum diagrammatic paradigm
that applies to arbitrary observables in any $\dagger$-SMC
\cite{CPaqPav}: \em classical systems are encoded as a single wire and
quantum systems as a double wire\em. The same applies to operations,
and $m_{\smallgraydot}$ and $m_{\smallgraydot}^\dagger$ allow passage
between these types. 

Note that the classical data will `remember' to which observable it
relates,  cf.~the encoding $\sum_i p_{i} \ket{x_i}$.  This is
physically meaningful since, for example, when one measures position
the resulting value will carry specification of the length unit in
which it is expressed.  If one wishes to avoid interconversion of this
`classical data with memory', one could fix one observable, and
unitarily transform the quantum data before measuring. Indeed, if  
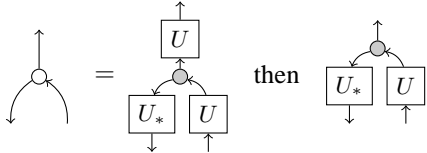
\[
\beginpgfgraphicnamed{Obs_transf}
\InputIfFileExists{Obs_transf.tikz}{}{\input{./figures/Obs_transf.tikz}}
\endpgfgraphicnamed \quad \mbox{then} \quad %
\beginpgfgraphicnamed{Obs_transf2}
\InputIfFileExists{Obs_transf2.tikz}{}{\input{./figures/Obs_transf2.tikz}}
\endpgfgraphicnamed
\]
measures the $\whiteobs$-observable but produces $\grayobs$-data. In \catFHilb, all observable structures are unitarily isomorphic, so any projective measurement can be obtained in this way. A particularly relevant example is when these unitaries are phases with respect the another observable structure $\whiteobs$.

\begin{equation}\label{eq:alphameasurement}
\graymeas^\alpha := %
\beginpgfgraphicnamed{Obs_transf3}
\InputIfFileExists{Obs_transf3.tikz}{}{\input{./figures/Obs_transf3.tikz}}
\endpgfgraphicnamed
\end{equation}

When $\whiteobs$ is induced by the Pauli spin-$Z$ observable and
$\grayobs$ by the Pauli spin-$X$ observable, then $\graymeas =
\graymeas^0$ is an $X$ measurement and $\graymeas^{\pi/2}$ is a $Y$
measurement. Note however, that both produce Born vectors of outcome
probabilities with respect to the $\graydot$ basis. This will be
useful in the sections to come. 

% is  the measurement with eigenstates $\left\{\left(\begin{array}{c}1\\ \!\!e^{i\alpha}\!\!\! \end{array}\right), \left(\begin{array}{c} 1\\ \!\!e^{i(\alpha+\pi)}\!\!\!\end{array}\right)\right\}$.  The phase group enables one to generalise this concept of a \em measurement against angle $\alpha$ \em to  observables in process theories.

\section{Strong Complementarity}\label{Sec:complementarity}

%%%% DISCUSSION OF ANTIPODE HERE
% 1. always self adjoint
% 2. homomorphism of monoid / comonoid structure
% what else?

\begin{definition}
A pair $(\whiteobs, \grayobs)$ of observables on the same object
$X$ is \em complementary \em iff:
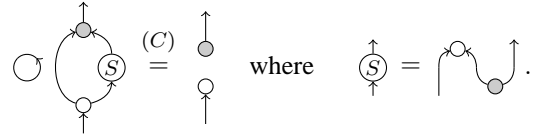
\[
\beginpgfgraphicnamed{antipode_hopf}
\InputIfFileExists{antipode_hopf.tikz}{}{\input{./figures/antipode_hopf.tikz}}
\endpgfgraphicnamed \quad\ \mbox{\rm where} \ \ \quad %
\beginpgfgraphicnamed{mub_antipode}
\InputIfFileExists{mub_antipode.tikz}{}{\input{./figures/mub_antipode.tikz}}
\endpgfgraphicnamed\ .
\]
\end{definition}
If at least one of the two observables has `enough classical points', this
equation holds if and only if the classical points of one observable
are `unbiased' (in sense of \cite{CD}) for the other observable.
Every observable in ${\bf FHilb}$ has enough classical points, hence
we reclaim the usual notion of quantum complementarity, and extend it
to a more general setting.

%  in particular to ${\bf Stab}$ and ${\bf
%   Spek}$.  For generalised observables for which at least one has
% `enough classical points' one can show that this equation holds if and
% only if classical points of one observable are `unbiased' (in sense of
% \cite{CD}) for the other observable, hence mimicking the intended
% operational meaning of the concept in ordinary quantum theory. 

% In the case that $S=1_X$ this equation simplifies to: 
% \begin{equation}\label{eq:simpleHopf}
%  \tikzfig{op_dir_hopf}\ ,
% \end{equation}
% that is, wires between complementary nodes cancel pairwise. 

\begin{definition}
A pair $(\whiteobs, \grayobs)$ of observables on the same object
$X$ is \em coherent \em iff: 
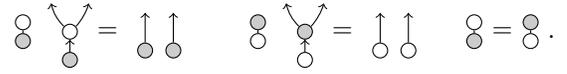
\[
\beginpgfgraphicnamed{coher}
\InputIfFileExists{coher.tikz}{}{\input{./figures/coher.tikz}}
\endpgfgraphicnamed\ .
\]
\end{definition}
In other words, $\whiteepsilon$ is proportional to a classical
point for $\grayobs$, and vice versa.
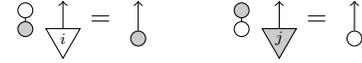
\[
\beginpgfgraphicnamed{classpoint-cohere}
\InputIfFileExists{classpoint-cohere.tikz}{}{\input{./figures/classpoint-cohere.tikz}}
\endpgfgraphicnamed
\]
We will assume that the scalar $%
\beginpgfgraphicnamed{scalar}
\InputIfFileExists{scalar.tikz}{}{\input{./figures/scalar.tikz}}
\endpgfgraphicnamed$ is always
cancelable.  

% From the first equation it easily follows that
% $\circl=\tikzfig{scalar}\, \tikzfig{scalar}$\,.\footnote{\bR Thir
%   requires coinciding compact structures right? One solution would be
%   to never consider the circle in this section but always a pair of
%   $\tikzfig{scalar}$'s. \e} 

\begin{proposition}\label{prop:cohere}\em
In ${\bf FHilb}$ if $O_{\!\smallwhitedot}$ and $O_{\!\smallgraydot}$ are self-adjoint operators correspoding to complementary observables, one can always choose a pair of coherent observable structures $(\whiteobs,\grayobs)$ whose classical points correspond to the eigenbases of $O_{\!\smallwhitedot}$ and $O_{\!\smallgraydot}$.
\end{proposition}

\begin{IEEEproof}
  (sketch) The eigenbasis of a non-degenerate self-adjoint operator is only determined up to global phases. For a pair of mutually unbiased bases, it is always possible to choose these phases such that coherence is satisfied.
\end{IEEEproof}

For this reason we will from now on assume that pairs of complementary observables are always coherent. 

\begin{definition}
A pair $(\whiteobs, \grayobs)$ of observables on the same object
$X$ is \em strongly complementary \em iff they are coherent and: 
\begin{equation}\label{eq:bialg}
\beginpgfgraphicnamed{bialg}
\InputIfFileExists{bialg.tikz}{}{\input{./figures/bialg.tikz}}
\endpgfgraphicnamed 
\end{equation} 
\end{definition}

Viewing one observable as monoid and the other as comonoid, the
properties of coherence and strong complementarity state that a
strongly complementary pair $(\whiteobs,\grayobs)$ form a
\emph{scaled bialgebra}; that is, the defining equations of a
bialgebra \cite{StreetBook} hold upto a scalar multiple.

The following results about the antipode for a strongly complementary pair were shown in~\cite{Aleks}.

\begin{lemma}\label{lem:antipode}
  Under the assumption that classical points are self-conjugate in their own colour, and provide `enough points', the antipode $S$ is self-adjoint, and is a Frobenius algebra endomorphism in both colours.
\end{lemma}

In fact we can go further.

\begin{theorem}\label{thm:SCimpliesC}\em
Strong complementarity $\Rightarrow$ complementarity.
\end{theorem}
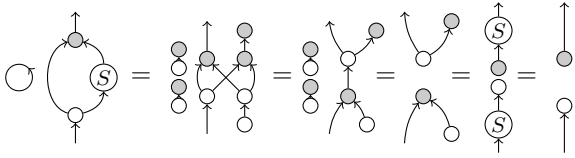
\begin{IEEEproof}\par
 \centerline{%
\beginpgfgraphicnamed{sc_hopf_pf}
\InputIfFileExists{sc_hopf_pf.tikz}{}{\input{./figures/sc_hopf_pf.tikz}}
\endpgfgraphicnamed}
\end{IEEEproof}

As a consequence, strongly complementary observables always form a \em 
scaled Hopf algebra\em.  Note that Theorem \ref{thm:SCimpliesC} relies
on the fact that both the monoid and the comonoid form a Frobenius
algebra; it is certainly not the case that every scaled bialgebra is a
Hopf algebra. 

The converse to Theorem~\ref{thm:SCimpliesC} does not hold:  it is
possible to find coherent complementary observables in ${\bf FHilb}$
which are not strongly complementary.  See \cite{CD2} for a
counterexample. 

\subsection{Strong Complementarity and Phase Groups} 

For  complementary observables, classical points of one
observable are always included in the phase group of the other
observable, up to a normalizing scalar. Strong complementarity strengthens
this property to inclusion as a subgroup.  
%The proofs of the following two results can be found in~\cite{CD}.
Let ${\cal K}_{\!\smallgraydot}$ be the set of classical points of $\grayobs$ multiplied by the scalar factor $%
\beginpgfgraphicnamed{scalar}
\InputIfFileExists{scalar.tikz}{}{\input{./figures/scalar.tikz}}
\endpgfgraphicnamed$.

\begin{theorem}\em\label{thm:subphasegroup} 
  Let $(\whiteobs, \grayobs)$ be strongly complementary
  observables and let $\grayobs$ have finitely many classical
  points. %, ${\cal K}_{\!\smallgraydot}$.  
  Then ${\cal
    K}_{\!\smallgraydot}$ forms a subgroup of the phase group
  $\Phi_{\!\smallwhitedot}$ of $\whiteobs$.  The converse also holds
  when $\whiteobs$ has `enough classical points'.
\end{theorem}
\begin{IEEEproof}
By strong complementarity it straightforwardly follows that, up to a scalar, $\whitemu$ applied to two classical points of $\grayobs$ yields again a classical point of $\grayobs$:
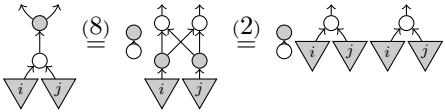
\[
\beginpgfgraphicnamed{bialgclosedness}
\InputIfFileExists{bialgclosedness.tikz}{}{\input{./figures/bialgclosedness.tikz}}
\endpgfgraphicnamed
\]
The unit of $\Phi_{\!\smallwhitedot}$ is, up to a scalar, also a classical point of $\grayobs$ by coherence. Hence,  ${\cal K}_{\!\smallgraydot}$ is a submonoid of $\Phi_{\!\smallwhitedot}$ and any finite submonoid is a subgroup.  The converse follows from:  
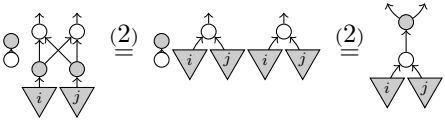
\[
\beginpgfgraphicnamed{bialgclosedness2}
\InputIfFileExists{bialgclosedness2.tikz}{}{\input{./figures/bialgclosedness2.tikz}}
\endpgfgraphicnamed
\]
together with the `enough classical points' assumption.
\end{IEEEproof}

Recall that the exponent of a group $G$ is the maximum order of any element of that group: $\textrm{exp}(G) = \textrm{max}\{ |g| : g \in G \}$.

\begin{corollary}\label{cor:orderHopf}\em
For any pair of strongly complementary observables, let $k = \textrm{\rm exp}({\cal K}_{\!\smallgraydot})$. Then, assuming $\grayobs$ has `enough classical points':
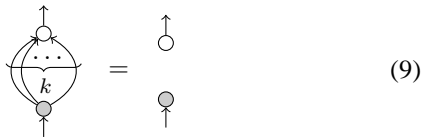
\begin{equation}\label{eq:orderHopf}
\beginpgfgraphicnamed{orderHopf}
\InputIfFileExists{orderHopf.tikz}{}{\input{./figures/orderHopf.tikz}}
\endpgfgraphicnamed
\end{equation}
\end{corollary}
\begin{IEEEproof}
In a finite abelian group, the order of any element divides $\textrm{\rm exp}({\cal K}_{\!\smallgraydot})$. The result then follows by:
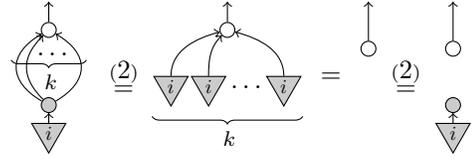
\[
\beginpgfgraphicnamed{orderHopfproof1}
\InputIfFileExists{orderHopfproof1.tikz}{}{\input{./figures/orderHopfproof1.tikz}}
\endpgfgraphicnamed
\] 
 together with the `enough classical points' assumption.
\end{IEEEproof}

\begin{proposition}\label{prop:alterdef}\em
For a pair of strongly  complementary observables
$\raisebox{1mm}{%
\beginpgfgraphicnamed{classicalpointaction}
\InputIfFileExists{classicalpointaction.tikz}{}{\input{./figures/classicalpointaction.tikz}}
\endpgfgraphicnamed}$ is a $\grayobs$-homomorphism for all
$%
\beginpgfgraphicnamed{classicalpointwhite}
\InputIfFileExists{classicalpointwhite.tikz}{}{\input{./figures/classicalpointwhite.tikz}}
\endpgfgraphicnamed\in{\cal
  K}_{\!\smallgraydot}$. Conversely,  
this property defines strong complementarity provided $\whitedelta$
has `enough classical points'. 
\end{proposition}
\begin{IEEEproof}
Similar to the proof of Thm.~\ref{thm:subphasegroup}.
\end{IEEEproof}

% Let $\Owg$ denote the quotient group $\whitePhi / \grayK$; ${\cal O}$ 
% stands for `observable' as the next proposition justifies.

% \begin{proposition}\label{prop:cosetsmeas}\em
%   For a pair of strongly complementary observables, for each coset
%   $O\in \Owg$ there is an observable $\graydelta^\alpha$ for which the
%   elements of $O$ are classical points.
% \end{proposition}
% \begin{IEEEproof} 
%   For $\alpha\in O$ set
%   $\graydelta^\alpha:=(\overline{\alpha}\otimes\overline{\alpha})\circ\graydelta\circ\alpha$.
%   By Prop.~\ref{prop:alterdef} it follows that $\graydelta^\alpha$
%   does not depend on the choice of $\alpha\in O$\footnote{\bR Put
%     graphical proof of this?\e} and classicality of $\alpha'=
%   \alpha\circ\tikzfig{classicalpointwhite} \in O$ follows
%   straightforwardly from classicality of
%   $\tikzfig{classicalpointwhite}$ for $\graydelta$.
% \footnote{\bR Put graphical proof of this? - ANY proof would be nice,
%   seems obvioulsy false for some interpretations of
%   ``$\graydelta^\alpha$''\e} 
% \end{IEEEproof}

\subsection{Classification of Strong Complementarity in ${\bf FHilb}$} 

%Theorem \ref{thm:subpahsegroup} straightforwardly yields  the classification of strongly complementary observables  in ${\bf FHilb}$. %, a result that was first observed in one of the authors' DPhil thesis \cite{Aleks}.

\begin{corollary}\em\label{col:classification}
Every pair of strongly complementary observables in ${\bf FHilb}$ is
of the following form: 
\[
\left\{\begin{array}{cl}
\!\!\graydelta&\!\!\!\!\!\!::|g\rangle\mapsto |g\rangle\otimes |g\rangle\vspace{1mm}\\
\!\! \grayepsilon&\!\!\!\!\!\!::|g\rangle\mapsto 1
\end{array}\right.
\quad
\left\{\begin{array}{cl}
\!\!\whitedelta^\dagger&\!\!\!\!\!\!:: |g\rangle\otimes |h\rangle\mapsto {1\over\sqrt{D}} |g + h\rangle\vspace{1mm}\\
\!\! \whiteepsilon^\dagger&\!\!\!\!\!\!::1\mapsto \sqrt{D} |0\rangle
\end{array}\right.
\]
where $(G =\{g, h, \ldots\}, +, 0)$ is a finite Abelian
group. Conversely, each such pair is always strongly complementary.  
\end{corollary}
\begin{IEEEproof}
By Theorem \ref{thm:subphasegroup} it follows that the classical
points of one observable (here  $\grayobs$) form a group for the
multiplication of the other observable (here $\whitedelta^\dagger$),
and in ${\bf FHilb}$ this characterises strong complementarity.   
\end{IEEEproof}
%Note that $G$ in this theorem is just \grayK as a subgroup of $\whitePhi$.

%%%% this has nothing to do with MUBs... see note below.
One of the longest-standing open problems in quantum information is
the characterisation of the number of pairwise complementary
observables in a Hilbert space of dimension $D$.  In all known cases
this is $D+1$, and the smallest unknown case is $D=6$.  We now show
that in the case of strong complementarity this number is always $2$
for $D\geq 2$.

%\footnote{\bR Ross, Aleks; you know of anything more that
%   fits in this section?\e} 

\begin{theorem}\em 
  In a Hilbert space with $D\geq 2$ the largest set of pairwise
  strongly complementary observables has size at most $2$.
\end{theorem}
%%%%% NB - as a fact about FDhilb, strong complementarity is not
%%%%% needed. If A and B are complimentary (i.e mub) take A to be the
%%%%% std basis.  Coherence now forces b_1 to (1,1,..,1).  If C is
%%%%% also coh. comp. to A then B and C must share a vector.
\begin{IEEEproof}
Assume that both $(\whiteobs, \grayobs)$ and  $(\whiteobs,
\blackobs)$ are strongly complementary pairs.  By coherence
$\grayunit$ and $\unit$ must be proportional to classical points of
$\whiteobs$. If $(\grayobs, \blackobs)$ were to be strongly
complementary observables, it is easily shown that
$%
\beginpgfgraphicnamed{innerprod}
\InputIfFileExists{innerprod.tikz}{}{\input{./figures/innerprod.tikz}}
\endpgfgraphicnamed\not= 0$ so $\grayunit$ and $\unit$ are
proportional to the same classical point.  Hence, up to a non-zero
scalar: 
\[
\beginpgfgraphicnamed{two_strong_compl_pfBIS}
\InputIfFileExists{two_strong_compl_pfBIS.tikz}{}{\input{./figures/two_strong_compl_pfBIS.tikz}}
\endpgfgraphicnamed
\]
i.e.~the identity has rank 1, which fails for $D\geq2$. By Corollary \ref{col:classification} a strongly complementary pair  exists for  any $D\geq2$.
\end{IEEEproof}

\section{Diagrammatic Computation of GHZ Measurement Outcome Distributions}\label{sec:corcomp}

In order to present a graphical Mermin/GHZ style arguement, we show
how to compute measurement outcomes for an $n$-party GHZ state
graphically. This computation relies crucially on the following
corollary, which follows from strong complementarity via a standard
theorem about bialgebras. 

\begin{corollary}\label{cor:genbialg}
The following equation holds for any connected bipartite graph with directions as shown.
\begin{equation}\label{eq:bialgarrows}
\beginpgfgraphicnamed{directed_bialg_cor}
\InputIfFileExists{directed_bialg_cor.tikz}{}{\input{./figures/directed_bialg_cor.tikz}}
\endpgfgraphicnamed
\end{equation}
\end{corollary}
The proof is given in Appendix \ref{sec:some-categ-backgr}.

We compute the classical probability distributions (= $\grayobs$-data) for $n$ measurements against arbitrary phases $\alpha_i\in \Phi_{\!\smallwhitedot}$  on $n$ systems of any type in a  generalised $GHZ^n_{\!\smallwhitedot\!}$-state:
\[ %
\beginpgfgraphicnamed{CorrelationComp1}
\InputIfFileExists{CorrelationComp1.tikz}{}{\input{./figures/CorrelationComp1.tikz}}
\endpgfgraphicnamed
   \stackrel{(\ref{eq:decspidercomp})}{=}
\beginpgfgraphicnamed{CorrelationComp2}
\InputIfFileExists{CorrelationComp2.tikz}{}{\input{./figures/CorrelationComp2.tikz}}
\endpgfgraphicnamed = (*) \]

Applying Corollary \ref{cor:genbialg}, we note that this is a probability distribution followed by a $\whitedot$-copy.
\begin{equation}\label{eq:correlationspider}
(*) = %
\beginpgfgraphicnamed{CorrelationComp3}
\InputIfFileExists{CorrelationComp3.tikz}{}{\input{./figures/CorrelationComp3.tikz}}
\endpgfgraphicnamed =: %
\beginpgfgraphicnamed{angle_dist}
\InputIfFileExists{angle_dist.tikz}{}{\input{./figures/angle_dist.tikz}}
\endpgfgraphicnamed 
\end{equation}

The following is an immediate consequence.
\begin{theorem}\label{thm:symmetriccorr}\em
When measuring each system of a $GHZ^n_A$-state against an arbitrary
angle then the resulting classical probability distribution over
outcomes is symmetric.  
\end{theorem}

% \begin{proposition}\label{prop:sublementarity}\em 
% If $\graydelta$ has `enough classical points' and if  
% \bR  $\raisebox{1.5mm}{\tikzfig{classicalpointwhiteinner}}=\delta_{ij}$ (needs detail in  most economical and multiplicative manner possible) \e\footnote{\bR Note that this can be seen as an assumption on classical data, i.e.~at the level of mixing where additive structure is more easily conceptually justified; this might need a discussion earlier on in the paper.\e} then $\alpvec=\tikzfig{classicalpointwhite}$ implies 
% $\sum_i\alpha_i=\tikzfig{classicalpointwhite}$\,.
% \end{proposition}
% \begin{IEEEproof}
% \bR ... plugging ... \e 
% \end{IEEEproof}
% Hence, if the assumptions of Prop.~\ref{prop:sublementarity} hold, $\alpvec:=\sum_i\alpha_i$.

\begin{theorem}\em
The classical probability distributions for
$\graymeas^{\alpha_1}\otimes\ldots\otimes
\graymeas^{\alpha_n}$-measurements on a $GHZ^n_A$-state is: 
\begin{itemize}
\item uncorrelated if $\roundket{\sum \alpha_i}$ is a classical point for $\whiteobs$ and,
\item parity-correlated if $\roundket{\sum \alpha_i}$ is a classical point $i$ for $\grayobs$ (i.e. contains precisely those outcomes $i_1 \otimes \ldots \otimes i_n$ such that the sum of group elements $\sum i_k$ is equal to $i$).
\end{itemize}
\end{theorem}

\begin{example}
We can compute the outcome
distributions for $XXX$,  $XYY$, $YXY$, and $YYX$ measurements on three qubits in a GHZ-state using the technique described above. First, outcome distribution $\roundket{A}$ for $XXX$:
\[
\beginpgfgraphicnamed{XXX_corrs}
\InputIfFileExists{XXX_corrs.tikz}{}{\input{./figures/XXX_corrs.tikz}}
\endpgfgraphicnamed
\]
Next, we compute outcome distribution $\roundket{B_1}$ for $XYY$:
\[
\beginpgfgraphicnamed{XYY_corrs}
\InputIfFileExists{XYY_corrs.tikz}{}{\input{./figures/XYY_corrs.tikz}}
\endpgfgraphicnamed
\]
Clearly, the other two cases, $YXY$ and $YYX$, will give the same result. We therefore set $\roundket{B_1} = \roundket{B_2} = \roundket{B_3}$.
\end{example}

\section{Mermin's Non-locality Argument in CQM}\label{sec:GHZMerminAbs}

For a particular $n$-party  state $|\Psi\rangle$ in some theory,
a \em local hidden variable (LHV) \em  model  for that state consists
of: 
\begin{itemize}
\item a family of hidden states $|\lambda\rangle$, each of  which
  assigns for any measurement on each  subsystem a definite outcome,  
\item and, a probability distribution on these hidden states,
\end{itemize}
which simulates the probabilities of that theory. We say that a theory
is \em local \em if each state admits a LHV model.

The Mermin argument is an important thought experiment which rules out
the possibility that the predictions of quantum mechanics could be
explained by LHV models~\cite{Mermin}.  Unlike  Bell's argument, which
merely shows that there exist quantum mechanical states whose outcome
\textit{probabilities} are inconsistent with locality assumptions, the
Mermin argument demonstrates how there exists a quantum state whose
outcome \textit{possibilities} are inconsistent with locality.   
 
Consider three  systems and four possible  (compound)  measurement
settings, consisting of the \textit{control} $XXX$, and three
\textit{variations} $XYY$, $YXY$, and $YYX$.  Let:
\ctikzfig{global_hidden_state}
be a Born vector for $\grayobs$ which represents the probability
distribution on possible outcomes for each of these settings.  This
Born vector is a probability distribution over `hidden states' each of
which specifies the outcomes of all three systems for each of the four
measurement settings, for example: 
\[ \ket\lambda =  |\ 
  \underbrace{+--}_{XXX}\  \underbrace{+++}_{XYY}\ 
  \underbrace{--+}_{YXY}\  \underbrace{-+-}_{YYX}\ 
 \rangle \]
yields outcomes $(+1,-1,-1)$ when $XXX$ is measured, outcomes
$(+1,+1,+1)$ when $XYY$ is measured. 

Assume now that this Born vector arises from an underlying LHV model
(L).  A hidden state of the LHV model stores one measurement outcome
for each setting on each system: 
\[ \ket{\lambda'} =  |\ 
  \underbrace{\overbrace{+}^X\overbrace{-}^Y}_{\textit{\footnotesize system~1}}\  
  \underbrace{\overbrace{-}^X\overbrace{+}^Y}_{\textit{\footnotesize system~2}}\ 
  \underbrace{\overbrace{-}^X\overbrace{+}^Y}_{\textit{\footnotesize system~3}}\
 \rangle \]

%In other words, the settings and outcomes on any of the three qubits do not affect the outcomes on the other two. This is akin to the local hidden variable states described by Mermin. As before, 
We can represent a probability distribution over such hidden states as
a Born vector $\roundket{\Lambda'}$, from which we generate
$\roundket{\Lambda}$  via copy maps, which put a hidden state
$\ket{\lambda'}$ into the form $\ket\lambda$: 
\begin{center}
\beginpgfgraphicnamed{global_hidden_state}
\InputIfFileExists{global_hidden_state.tikz}{}{\input{./figures/global_hidden_state.tikz}}
\endpgfgraphicnamed
$\overset{(L)}{=}$
\beginpgfgraphicnamed{local_hidden_state}
\InputIfFileExists{local_hidden_state.tikz}{}{\input{./figures/local_hidden_state.tikz}}
\endpgfgraphicnamed
\end{center}
For example, the leftmost wire coming from $\roundket{\Lambda'}$
represents the outcome when the first qubit of each hidden state is
measured in $X$. This outcome is copied and sent to the first qubit for
setting $XXX$ and the first qubit for setting $XYY$.
%The second wire represents the outcome when the first qubit is measured in $Y$. This is copied and sent to the first qubit for setting $YXY$ and the first qubit for setting $YYX$. The outcomes for the other two qubits represented by $\roundket{\Lambda'}$ are handled similarly. 
%If $\roundket{\Lambda}$ is of the form above, it is said to satisfy \textit{locality}.

We now investigate whether measurement of $XXX$, $XYY$, $YXY$ and
$YYX$ on the qubits in a GHZ state is \textit{consistent} (C) with
such a LHV model.
%Each hidden state in $\roundket{\Lambda}$ gives outcomes for all four measurement setups simultaneously. 
Although it is impossible to perform all four measurement setups
simultaneously on the three qubits in the GHZ state, it should at
least be true that any possible hidden state in $\roundket{\Lambda}$
must not be ruled out by the combined experimental data.

Since we are assuming that $\roundket{A} \otimes \roundket{B_1}
\otimes \roundket{B_2} \otimes \roundket{B_3}$ is the probability
distribution yielded by sampling $\roundket\Lambda$ independently for
each measurement setup, it is a course-graining of $\roundket\Lambda$
itself. In other words, the support of $\roundket\Lambda$ must be
contained in the support of $\roundket{A} \otimes \roundket{B_1}
\otimes \roundket{B_2} \otimes \roundket{B_3}$.\footnote{Note that
  this is
%in the spirit of the usual Mermin argument, as it is 
  a statement about \textit{possibilities} rather than
  \textit{probabilities}.}  In particular, if we can apply some
function on hidden states in $\roundket{A} \otimes \roundket{B_1}
\otimes \roundket{B_2} \otimes \roundket{B_3}$ that yields a definite
outcome, that function must yield the same outcome in
$\roundket\Lambda$. Consider the function:

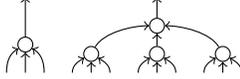
\ctikzfig{constant_function}
\noindent
This function computes the parity (i.e. the $\mathbb Z_2$-sum) of
outcomes for each of the four experiments. It then furthermore
computes the \textit{overall} parity of outcomes for the 3 variation
experiments. We can straightforwardly show this yields a constant
outcome on $\roundket{A} \otimes \roundket{B_1} \otimes \roundket{B_2}
\otimes \roundket{B_3}$: 

% TODO: fix EQ number in this vvvv
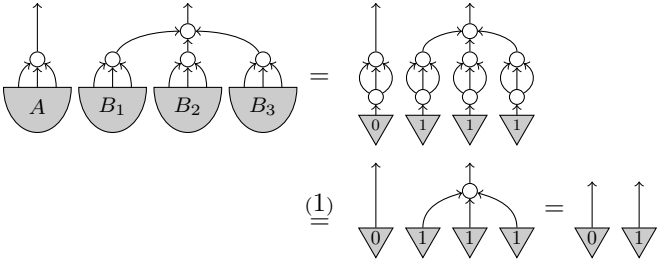
\ctikzfig{parity_outcomes_corrected}

\noindent
%The last equation follows because $\whitemu$ forms a group algebra over the classical points of $\grayobs$. In two dimensions, this must be $\mathbb Z_2$. From this, we conclude:

The last equation follows because by Theorem~\ref{thm:subphasegroup} the classical points for $\grayobs$ form a subgroup of the phase group $\whitePhi$. Since there are two classical points for $\grayobs$, this group must be $\mathbb Z_2$.

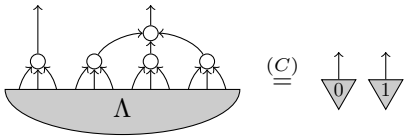
\ctikzfig{hidden_var_outcomes} 

We can then use equations $(C)$ and $(L)$ to derive a
contradiction. First, substitute $(L)$ into the LHS of $(C)$:

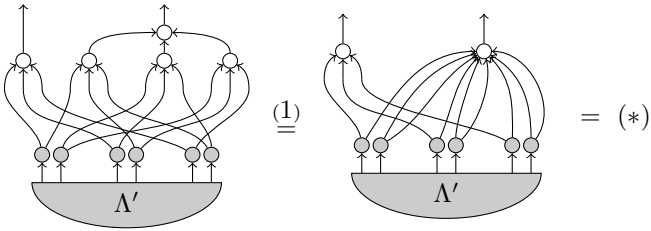
\ctikzfig{mermin_pf1}
\noindent
Since the group associated with $\whitedot$ is $\mathbb Z_2$,
Cor.~\ref{cor:orderHopf} applies for $k = 2$. Thus, we can delete
pairs of parallel edges connecting dots of different colours. Then,

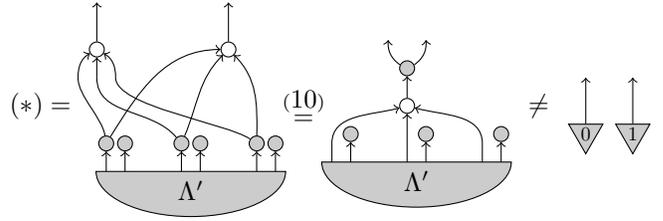
\ctikzfig{mermin_pf2}
\noindent
The final inequation follows from the distinctness of classical points
and the cocommutativity of $\graycomult$. Note that the symmetry from
the LHS arises from the fact that after the application of
Cor.~\ref{cor:orderHopf}, only the X measurements in the variations
contribute to the output of the function.
 
\section{Generalised Non-locality Arguments}\label{sec:Mermin_general}  

To summarise, the conflict exposed in the previous section arises from
the fact that experimental data, here from quantum theory, excludes a
property imposed on the hidden states by locality.  That property was
symmetry. We now generalise this.

First we consider more general lists of measurement settings,
represented again by a Born vector for $\grayobs$ generated from a
LHV model:
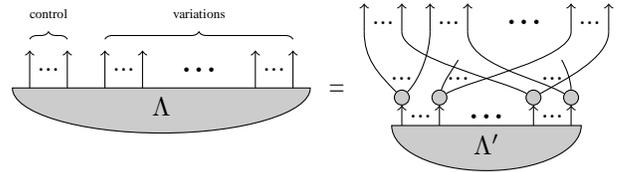
\begin{center}
\beginpgfgraphicnamed{global_hidden_state_bis}
\InputIfFileExists{global_hidden_state_bis.tikz}{}{\input{./figures/global_hidden_state_bis.tikz}}
\endpgfgraphicnamed = %
\beginpgfgraphicnamed{local_hidden_state_bis}
\InputIfFileExists{local_hidden_state_bis.tikz}{}{\input{./figures/local_hidden_state_bis.tikz}}
\endpgfgraphicnamed
\end{center}
where we require, that via application of  Cor.~\ref{cor:orderHopf}, we again obtain a symmetric expression:
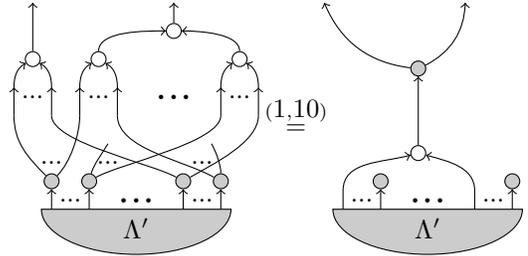
\ctikzfig{mermin_pf1_bis}

\noindent
This requires that for the variations, the multiplicity of occurrence
of a particular measurement on the same system has to be a multiple of
$exp({\cal K}_{\!\smallgraydot})$.
  
Contradicting  this symmetry  requires an inequation between classical points:
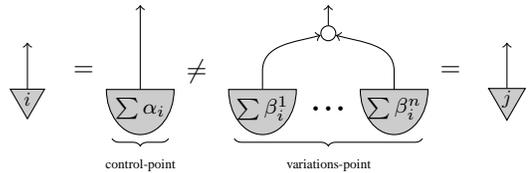
\ctikzfig{parity_outcomes_tris}
  
 \begin{theorem}\em 
The above scenario provides a generalised Mermin non-locality  argument whenever:
\begin{enumerate}
\item the multiplicity of occurrence of measurements on the same system in the variations is  a multiple of $exp({\cal K}_{\!\smallgraydot})$,  
\item and, the control-point and the variations-point are distinct classical points.
\end{enumerate}
  \end{theorem}

This theorem yields a wide variety of generalised Mermin-type non-locality  arguments, and also  characterises situations where such an argument fails to hold.
  
% We illustrate this by means of two examples.
  
%   \begin{corollary} 
% If $\whitePhi={\cal  K}_{\!\smallgraydot}\times  (\whitePhi/{\cal  K}_{\!\smallgraydot})$ then no generalised Mermin non-locality  argument is possible.
%  \end{corollary} 
%  \begin{IEEEproof}
% \bR include proof.\e
% \end{IEEEproof}

%  An example of this is Spekken's toy theory \cite{Spekkens} which can be cast as a  category {\bf Spek} \cite{Spek, CES}. In this case we have  $ \whitePhi= \mathbb{Z}_2\times \mathbb{Z}_2={\cal  K}_{\!\smallgraydot}\times  (\whitePhi/ {\cal  K}_{\!\smallgraydot})$. 
 
%   \begin{corollary} 
%  Taking the control setting and the $i$-the of $exp({\cal  K}_{\!\smallgraydot})+1$ variation settings respectively to be
%  \[
%  \overbrace{
%  \graymeas^{0}\otimes\ldots\otimes \graymeas^{0}}^{exp({\cal  K}_{\!\smallgraydot})+1  
%  }
%  \quad
%   \overbrace{
%   \graymeas^{\beta}\otimes\ldots\otimes\graymeas^{\beta}\otimes\underbrace{\graymeas^{0}}_i\otimes 
% \graymeas^{\beta}\otimes\ldots\otimes\graymeas^{\beta} 
%   }^{exp({\cal  K}_{\!\smallgraydot})+1}
%   \]
%   where $\beta$ is chosen such that $exp({\cal  K}_{\!\smallgraydot})\cdot(exp({\cal  K}_{\!\smallgraydot})+1)\cdot\beta$ a classical point for $\grayobs$ different from the unit of $\whitePhi$,  yields a generalised  non-locality  argument.
%   \end{corollary} 
%    \begin{IEEEproof}
% \bR include proof.\e
% \end{IEEEproof}

%  Mermin's non-locality argument is an instance of this, namely the case that $exp({\cal  K}_{\!\smallgraydot})=2$ and $\beta={\pi\over 2}$ for qubits.

\section{Necessity of Strong Complementarity}\label{sec:converse}        

In this section, we make an argument that the assumptions of the Mermin argument necessitate the use of strongly complementary observables. Consider a fixed 3-party GHZ-state $(\whitedot)_0^3$. We will attempt to construct a Mermin argument by measuring all three systems with respect to some observables $O_1$, $O_2$, $O_3$. We make three assumptions about $O_i$.

\begin{enumerate}
  \item \textbf{Phase-related:} All three $O_i$ are within a $\whiteobs$-phase of some fixed observable structure $\grayobs$.
  \item \textbf{Coherence:} $\grayobs$ is coherent with respect to $\whiteobs$.
  \item \textbf{Sharpness:} After performing two of the three measurements on the GHZ-state, the remaining system is in an eigenstate of the third measurement.
\end{enumerate}

The first condition is satisfied by the usual Mermin argument, and can be seen as requiring that measurements differ in the `maximally non-local' manner, since $\whiteobs$-phases can pass freely through the GHZ state. In $\catFHilb$, when $\grayobs$ is already complementary, we can choose it to be coherent by \ref{prop:cohere}. The third assumption was highlighted by Mermin in~\cite{Mermin} as an important aspect of the experimental setup.

% Note that this is indeed a necessary condition for a Mermin argument. In order to have a classical point:
% \ctikzfig{parity_outcomes_quad}

% \noindent
% the outcome of the third measurement must fixed by the outcomes of the previous two measurements, since otherwise we would obtain an unsharp probability distribution.   It suffices to assume that there exists at least one set of angles $\alpha_1, \alpha_2, \alpha_3$ for which the above stated requirement is satisfied.

The map %
\beginpgfgraphicnamed{decoh}
\InputIfFileExists{decoh.tikz}{}{\input{./figures/decoh.tikz}}
\endpgfgraphicnamed is called the \textit{decoherence map} for $\grayobs$. It projects from the space of all quantum mixed states to the the space of classical mixtures of eigenstates of $\grayobs$. To assert sharpness, we require that, once two of the three systems are measured, the third is invariant under this map:
\begin{equation}\label{eq:ConverseProof1}
\beginpgfgraphicnamed{ConverseProof1}
\InputIfFileExists{ConverseProof1.tikz}{}{\input{./figures/ConverseProof1.tikz}}
\endpgfgraphicnamed
\end{equation}
Plugging the unit of $\whiteobs$ in the 2nd system both for LHS and RHS, and using coherence we obtain:
\begin{equation}\label{eq:ConverseProof2}
\beginpgfgraphicnamed{ConverseProof2}
\InputIfFileExists{ConverseProof2.tikz}{}{\input{./figures/ConverseProof2.tikz}}
\endpgfgraphicnamed 
\end{equation}
and  by exploiting symmetry we have: 
\begin{equation}\label{eq:ConverseProof3}
\beginpgfgraphicnamed{ConverseProof3}
\InputIfFileExists{ConverseProof3.tikz}{}{\input{./figures/ConverseProof3.tikz}}
\endpgfgraphicnamed
\end{equation}
Hence we obtain:
\[%\begin{equation}\label{eq:ConverseProof4} 
\beginpgfgraphicnamed{ConverseProof4}
\InputIfFileExists{ConverseProof4.tikz}{}{\input{./figures/ConverseProof4.tikz}}
\endpgfgraphicnamed
\]%\end{equation}
Since $\whitedelta^\dagger\circ(1_X\otimes \sum_i\alpha_i)$ is unitary it cancels.
 
\begin{proposition}\label{prop:bialgalt}\em
For a pair $(\whiteobs, \grayobs)$ of coherent observables on the same object the following equation implies (\ref{eq:bialg}):
\begin{equation}\label{eq:bialgalt}  
\beginpgfgraphicnamed{bialgaltbis}
\InputIfFileExists{bialgaltbis.tikz}{}{\input{./figures/bialgaltbis.tikz}}
\endpgfgraphicnamed
\end{equation} 
\end{proposition}

\begin{IEEEproof}
  See Appendix~\ref{sec:pf-bialgalt}.
\end{IEEEproof}

Thus, for any coherent pair of observables, the sharpness of the third outcome necessitates strong complementarity.

% \begin{IEEEproof} 
%   First we show that \eqref{eq:bialg} implies \eqref{eq:bialgalt}:
%   \[
%   \tikzfig{bialgaltbis-proof-i}
%   \]
%   where the first and last equalities were via \eqref{eq:bialg}.  Now for the
%   converse we have:
%   \begin{equation}\label{eq:bialgtbis-proof-i}
%     \tikzfig{bialgaltbis-proof-ii}
%   \end{equation}
%   where the second equation uses \eqref{eq:bialgalt}.  Now using
%   \eqref{eq:bialgtbis-proof-i} and then \eqref{eq:bialgalt} again we
%   have the result:
%   \[
%   \tikzfig{bialgaltbis-proof-iii}
%   \]
% \end{IEEEproof}

\section{Conclusion and Outlook}\label{sec:outlook} 

We cast Mermin's non-locality argument within CQM. This enabled us to substantially generalise this result to multiple parties, arbitrary angles and systems of arbitrary dimension. 

The tools of CQM made most computations very easy as compared to the methods of standard quantum theory.   We only rely on  two simple rules: (i) contraction of labeled nodes (cf.~(\ref{eq:decspidercomp})); (ii) commutation of the multiplications and comultiplications associated with distinctly colored nodes (cf.~(\ref{eq:bialgarrows})).

In the graphical language, the manner in which the phases associated to the measurements interact through the GHZ-state effectively \em depicts non-locality\em. 

The concepts required for  reproducing Mermin-style non-locality arguments as well as the derivations crucially relied on the newly introduced notion of strong complementarity.  We provided an operational interpretation for strong complementarity and classified strongly  complementary observables in the case of quantum theory.

Our analysis also provides new insights in Mermin's argument. For example, while we crucially rely on `strong' complementarity between the observable that characterises  the GHZ state and the one for which the classical points form a subgroup of the corresponding phase group, complementarity between the X and Y obsevables is not at all essential when passing to the general case. 

The results in this paper moreover open the door for a research program that concentrates on studying general process theories, in order to better understand what is so peculiar about the correlations encountered in quantum theory.  CQM indeed provides an ideal arena for relating key concepts of quantum theory, and investigating in which manner they survive when passing to more general process theories.

The results in this paper may also have more direct applications to quantum computing. In particular, quantum secret sharing (QSS) \cite{Hillery} is a protocol which very closely resembles Mermin's non-locality argument. It relies on $X$ or $Y$ measurements on a qubit in a  GHZ state, and the resulting correlations.  Clearly the results in this paper would enable one to substantially generalise this schemes in a similar manner that we generalised Mermin's non-locality argument.  Also, \cite{Anne} contains initial investigations for a CQM-treatment of a wide range of quantum informatic protocols, which could be given a similar treatment and may lead to more generalisations of quantum communication schemes.

\appendices
\section{Additional background}\label{sec:appnedix-addit-backgr}

\subsection{Some categorical background}\label{sec:some-categ-backgr}
We assume that the reader is familiar with basics of category theory,
and it comfortable with the notion of symmetric monoidal category
(SMC).   Suitable introductions to the subject are
\cite{Abramsky:2009fx,BobEric2011cats}.  We assume throughout that all
our monoidal categories are strict, i.e.~the morphisms $\alpha$,
$\rho$, and $\lambda$ are  identities.

\begin{definition}\label{def:dag-cat}
  A \emph{dagger category} is a category \catC equipped with a \emph{dagger
  functor} $\dag: \catCop \to \catC$ which is (i) involutive, and (ii)
  acts as the identity on objects.  
\end{definition}

\begin{definition}\label{def:unitary}
  In any dagger category, a morphism $f:A\to B$ is called
  \emph{unitary} if $f^\dagger\circ f = 1_A$ and $f\circ f^\dagger =
  1_B$.
\end{definition}

\begin{definition}\label{def:inner-product}
  Let $\psi,\phi:I \to A$ be points of some object.  We define the
  \emph{inner product} by $\braket{\psi}{\phi} := \psi^\dagger\circ \phi.$
\end{definition}
This definition coincides with usual one in the category of finite
dimensional  Hilbert spaces.

\begin{definition}\label{def:dsmc}
  A \emph{dagger symmetric monoidal category} ($\dagger$-SMC) is a
  symmetric monoidal category
  $(\catC,\otimes,I,\alpha,\rho,\lambda,\sigma)$ which is equipped
  with a strict monoidal dagger functor, such that the symmetry
  isomorphism  $\sigma$ is unitary\footnote{In a non-strict setting we
    would insist that the  other structure morphisms likewise be unitary.}.
\end{definition}

A primary method of this paper is the use of internal algebraic
structures defined inside monoidal categories.  We now introduce these,
one piece at a time.

\begin{definition}\label{def:monoid}
  A monoid in $\catC$ is a triple $(X, \mu : X \otimes X \rightarrow X, \eta : I \rightarrow X)$ such that the following diagrams commute:
  \begin{center}
    \begin{tikzpicture}
      \matrix (m) [cdiag] {
        X \otimes X \otimes X & X \otimes X \\
        X \otimes X & X \\
      };
      \path [arrs]
        (m-1-1) edge node {$X \otimes \mu$} (m-1-2)
        (m-2-1) edge node [swap] {$\mu$} (m-2-2)
        (m-1-1) edge node [swap] {$\mu \otimes X$} (m-2-1)
        (m-1-2) edge node {$\mu$} (m-2-2);
    \end{tikzpicture}
    
    \begin{tikzpicture}
      \matrix (m) [cdiag] {
        X &  \\
        X \otimes X & X \\
      };
      \path [arrs]
        (m-2-1) edge node [swap] {$\mu$} (m-2-2)
        (m-1-1) edge node [swap] {$\eta \otimes X$} (m-2-1)
        (m-1-1) edge node {$1_X$} (m-2-2);
    \end{tikzpicture}
    \quad
    \begin{tikzpicture}
      \matrix (m) [cdiag] {
        X & X \otimes X \\
          & X \\
      };
      \path [arrs]
        (m-1-1) edge node {$X \otimes \eta$} (m-1-2)
        (m-1-2) edge node {$\mu$} (m-2-2)
        (m-1-1) edge node [swap] {$1_X$} (m-2-2);
    \end{tikzpicture}
  \end{center}
  
  If \catC is symmetric then the monoid is \emph{commutative} if it further satisfies:
  \begin{center}
    \begin{tikzpicture}
      \matrix (m) [cdiag] {
        X\otimes X & X \\
        X\otimes X &   \\
      };
      \path [arrs]
        (m-1-1) edge node {$\mu$} (m-1-2)
        (m-2-1) edge node [swap] {$\mu$} (m-1-2)
        (m-1-1) edge node [swap] {$\sigma$} (m-2-1);
    \end{tikzpicture}
  \end{center}

  In diagrammatic notation we write $\mu = \whitemult$, $\eta = \whiteunit$; the
  equations can now be expressed as 
  \[
\beginpgfgraphicnamed{associativity}
\InputIfFileExists{associativity.tikz}{}{\input{./figures/associativity.tikz}}
\endpgfgraphicnamed
  \qquad  %
\beginpgfgraphicnamed{unit}
\InputIfFileExists{unit.tikz}{}{\input{./figures/unit.tikz}}
\endpgfgraphicnamed
  \qquad %
\beginpgfgraphicnamed{commutativity}
\InputIfFileExists{commutativity.tikz}{}{\input{./figures/commutativity.tikz}}
\endpgfgraphicnamed
  \]
\end{definition}

% \begin{definition}\label{def:monoid}
%   A \emph{monoid} in a monoidal category \catC consists of a triple
%   $(A,\mu,\eta)$, where the \emph{multiplication} $\mu:A\otimes A\to
%   A$ and the \emph{unit} $\eta:I\to A$ satisfy the following
%   equations:
%   \[
%   \begin{diagram}
%     A \otimes A \otimes A & \rTo^{\mu \otimes 1} & A \otimes A \\
%     \dTo<{1 \otimes \mu} & & \dTo>{\mu} \\
%     A \otimes A & \rTo_{\mu} & A
%   \end{diagram}
% \qquad
%   \begin{diagram}
%     A & \rTo^{\eta \otimes 1} & A \otimes A & \lTo^{1 \otimes \eta} &
%     A \\
%     & \rdTo_{1} & \dTo>{\mu} & \ldTo_{1} & \\
%     & A &
%   \end{diagram}
%   \]
%   If \catC is symmetric then the monoid is \emph{commutative} if it
%   further satisfies:
%   \[
%   \begin{diagram}
%     A \otimes A & \rTo^\sigma & A \otimes A \\
%     & \rdTo_{\mu} & \dTo>\mu \\
%     && A
%   \end{diagram}
%   \]
  % In diagrammatic notation we write $\mu = \whitemult$, $\eta = \whiteunit$; the
  % equations can now be expressed as 
  % \[
  % \tikzfig{associativity}
  % \qquad  \tikzfig{unit}
  % \qquad \tikzfig{commutativity}
  % \]
% \end{definition}

The dual to a monoid is a \emph{comonoid}.

\begin{definition}\label{def:comonoid}
  A \emph{comonoid} in a monoidal category  \catC consists of a triple
  $(X,\delta,\epsilon)$ which is a monoid in \catCop; i.e.~$\delta :
  X \to X \otimes X$ and $\epsilon : X \to I$ satisfy the equations of
  Definition~\ref{def:monoid} but in reverse, viz:
  \[
\beginpgfgraphicnamed{coassoc}
\InputIfFileExists{coassoc.tikz}{}{\input{./figures/coassoc.tikz}}
\endpgfgraphicnamed
  \qquad\qquad
\beginpgfgraphicnamed{counit}
\InputIfFileExists{counit.tikz}{}{\input{./figures/counit.tikz}}
\endpgfgraphicnamed
  \]
  A comonoid is \emph{cocommutative} if it satisfies:
  \[
\beginpgfgraphicnamed{cocomm}
\InputIfFileExists{cocomm.tikz}{}{\input{./figures/cocomm.tikz}}
\endpgfgraphicnamed.
  \]
\end{definition}

The basic example of a comonoid is in the category  of finite
dimensional Hilbert spaces, where for any basis $\{x_i\}_i$ of a space
$X$ we obtain a comonoid by defining `copying' and `erasing' maps:
\[
\delta : x_i \mapsto x_i \otimes x_i \qquad \epsilon : x_i \mapsto 1
\]
as the comultiplication and counit.

Notice that in a $\dagger$-SMC, if $(X,\delta,\epsilon)$ is a comonoid
then automatically $(X,\delta^\dagger,\epsilon^\dagger)$ is a monoid,
and vice versa.

\begin{definition}\label{def:comonoid-homomorphism}
  Given a comonoid $(\delta,\epsilon)$ on $X$, a \emph{comonoid
    homomorphism} is a map $f:X\to X$ such that 
\[
\delta \circ f = (f \otimes f) \circ \delta
\quad\text{ and }\quad
\epsilon \circ f = \epsilon\;.
\]
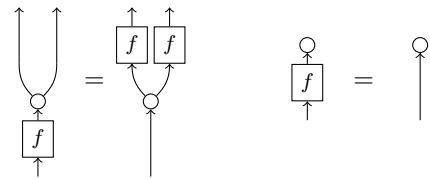
\[
\beginpgfgraphicnamed{comonoid-homo-i}
\InputIfFileExists{comonoid-homo-i.tikz}{}{\input{./figures/comonoid-homo-i.tikz}}
\endpgfgraphicnamed
\qquad\qquad
\beginpgfgraphicnamed{comonoid-homo-ii}
\InputIfFileExists{comonoid-homo-ii.tikz}{}{\input{./figures/comonoid-homo-ii.tikz}}
\endpgfgraphicnamed
\]
\end{definition}
A monoid homomorphism is defined dually.

The structures of greatest interest for this paper are algebras
containing both monoids and comonoids.

\begin{definition}\label{def:frobenius-alg}
  A \emph{commutative Frobenius algebra} is a 5-tuple
  $(X,\delta,\epsilon,\mu,\eta)$ where 
  \begin{enumerate}
  \item $(X,\delta,\epsilon)$ is a cocommutative comonoid;
  \item $(X,\mu,\eta)$ is a commutative monoid; and,
  \item $\delta$ and $\mu$ satisfy the following equations:
    \[
\beginpgfgraphicnamed{frob-condition}
\InputIfFileExists{frob-condition.tikz}{}{\input{./figures/frob-condition.tikz}}
\endpgfgraphicnamed
    \] 
  \end{enumerate}
  A Frobenius algebra is called \emph{special} if it additionally
  satisfies:
  \[
\beginpgfgraphicnamed{special-condition}
\InputIfFileExists{special-condition.tikz}{}{\input{./figures/special-condition.tikz}}
\endpgfgraphicnamed
  \]
\end{definition}

Let $\delta_n: X \to X^{\otimes n}$ be defined by the recursion
\[
\delta_0 := \epsilon 
\qquad 
\delta_{n+1} := (\delta_n \otimes 1_A) \circ \delta
\]
and define $\mu_m$ analogously.  Now we have the following important
theorem:
\begin{theorem}\label{thm:frob-alg-nf}
  Given a SCFA $(X,\delta,\epsilon,\mu,\eta)$ let $f:X^{\otimes m}\to
  X^{\otimes n}$ be a map constructed from $\delta$,$\epsilon$,$\mu$
  and $\eta$ whose graphical form is connected.  Then $f =
  \delta_n \circ \mu_m$.
\end{theorem}

This theorem is the basis of the `spider' notation introduced in
Proposition \ref{prop:spider} in the main text.

\begin{definition}\label{def:dscfa}
  In a $\dagger$-SMC, a \emph{dagger special commutative Frobenius
    algebra} ($\dagger$-SCFA) is a triple $(X,\delta,\epsilon)$ such
  that  $(X,\delta,\epsilon,\delta^\dagger,\epsilon^\dagger)$ is a
  special commutative Frobenius algebra.
\end{definition}

It has been shown \cite{CPV} that in ${\bf FHilb}$ \emph{every}
$\dagger$-SCFA arises from a comonoid defined by copying an
orthonormal basis as described above.  Since orthonormal bases define
non-degenerate quantum observables, $\dagger$-SCFAs are also called
\emph{observable structures}.

The other structures of interest, bialgebras and Hopf algebras also
arise via the interaction of a monoid and a comonoid.

\begin{definition}\label{def:bialgebra}
  A \emph{bialgebra} in a symmetric monoidal  category  is a 5-tuple
  $(X,\delta,\epsilon,\mu,\eta)$, where 
  \begin{enumerate}
  \item $(X,\delta,\epsilon)$ is a comonoid,
  \item $(X,\mu,\eta)$ is a monoid, and
  \item the  following equations are satisfied:
    
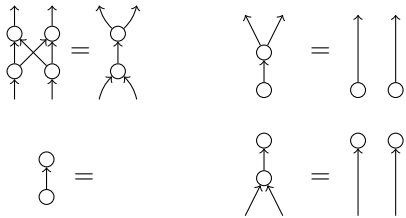
\[
    \begin{array}{ccc}
\beginpgfgraphicnamed{bialg-bialg}
\InputIfFileExists{bialg-bialg.tikz}{}{\input{./figures/bialg-bialg.tikz}}
\endpgfgraphicnamed &\qquad & %
\beginpgfgraphicnamed{bialg-copying}
\InputIfFileExists{bialg-copying.tikz}{}{\input{./figures/bialg-copying.tikz}}
\endpgfgraphicnamed\\
      \\
\beginpgfgraphicnamed{bialg-scalar}
\InputIfFileExists{bialg-scalar.tikz}{}{\input{./figures/bialg-scalar.tikz}}
\endpgfgraphicnamed\;\, &&       %
\beginpgfgraphicnamed{bialg-cocopying}
\InputIfFileExists{bialg-cocopying.tikz}{}{\input{./figures/bialg-cocopying.tikz}}
\endpgfgraphicnamed
    \end{array}
    \]
    where the empty diagram stands for $1_I$.
  \end{enumerate}
\end{definition}

In this paper, all the bialgebras we encounter will have the additional
property of being \emph{Hopf algebras}.

\begin{definition}\label{def:hopf-alg}
  A \emph{Hopf algebra} in a symmetric monoidal category is a 6-tuple
  $(X,\delta,\epsilon,\mu,\eta,s)$ where
  $(X,\delta,\epsilon,\mu,\eta)$ is a bialgebra, $s:X\to X$ is map
  called the \emph{antipode}, which altogether satisfy the equations:
  \[
\beginpgfgraphicnamed{hopf-law}
\InputIfFileExists{hopf-law.tikz}{}{\input{./figures/hopf-law.tikz}}
\endpgfgraphicnamed
  \]
\end{definition}

\begin{theorem}\label{thm:bialgebra-normal-form}
  Suppose  $(X,\delta,\epsilon,\mu,\eta)$ is a bialgebra and let
  $f:A^{\otimes m} \to A^{\otimes n}$ be a map constructed from
  $\delta$, $\epsilon$, $\mu$, and $\eta$.  Then $f$ factorises 
  $c\circ s \circ m$ where $c$ is constructed from the comonoid
  structure, $s$ is a permutation, and $m$ is constructed from the
  monoid structure.
  \[
  f \quad = \qquad %
\beginpgfgraphicnamed{bialg-nf}
\InputIfFileExists{bialg-nf.tikz}{}{\input{./figures/bialg-nf.tikz}}
\endpgfgraphicnamed
  \]
  Further, this factorisation is essentially unique.
\end{theorem}
We shall not present the full proof here; for that see \cite{Lack} and
refernces therein.  
Instead we treat only the case of interest, as needed for Corollary
\ref{cor:genbialg}.  Suppose that $f$ has the form $\delta_n \circ
\mu_{n'}$; we use induction on both $n$ and $n'$.  The base case $n = 0,
n'=0$, i.e.~$f = \epsilon\circ\eta$ is trivial.  If $n = 0$ and $n' > 0$
then $f$ has the form $\epsilon \circ \mu_{n'}$ by induction on $n'$
is equal to ${n'}$ copies of $\epsilon$.
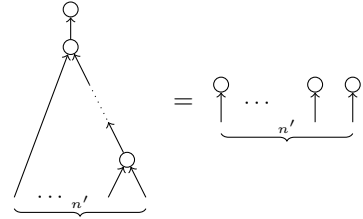
\[
\beginpgfgraphicnamed{bialg-nf-proof-i}
\InputIfFileExists{bialg-nf-proof-i.tikz}{}{\input{./figures/bialg-nf-proof-i.tikz}}
\endpgfgraphicnamed
\]
The case $n > 0, n' = 0$ is exactly dual.  Now suppose $n,n'>0$; we
have the situation pictured below.
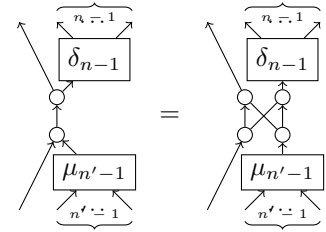
\[
\beginpgfgraphicnamed{bialg-nf-proof-ii}
\InputIfFileExists{bialg-nf-proof-ii.tikz}{}{\input{./figures/bialg-nf-proof-ii.tikz}}
\endpgfgraphicnamed
\]
Notice that we have $\delta_{n-1} \circ \mu$ in the top left;  by
induction we obtain the form show below:
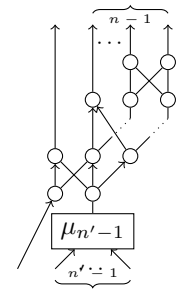
\[
\beginpgfgraphicnamed{bialg-nf-proof-iii}
\InputIfFileExists{bialg-nf-proof-iii.tikz}{}{\input{./figures/bialg-nf-proof-iii.tikz}}
\endpgfgraphicnamed
\]  
Similarly, we have $\delta \circ \mu_{n-1}$ on the bottom right and by
the same reasoning we can reduce $f$ to the required form.  Notice
that in both diagrams there is a path from every input to every
output.  This preservation of paths is the defining characteristic of
the normal form for bialgebras.

We can now apply the theorem to prove Corollary \ref{cor:genbialg}.

\begin{IEEEproof}[Proof of Corollary \ref{cor:genbialg}]
  For the theorem on bialgebras to apply, all of the edges need to be directed upward. For a strongly complementary observable, the edge direction between two different colours can be reversed by applying the dualiser. Then, we use the fact that $S$ is a Frobenius algebra endomorphism to move it down.
  \ctikzfig{directed_bialg_pf1}

  We apply Theorem \ref{thm:bialgebra-normal-form} and the spider theorem to complete the proof.
  \ctikzfig{directed_bialg_pf2}
\end{IEEEproof}

\subsection{Proof of Proposition \ref{prop:bialgalt}}\label{sec:pf-bialgalt}

We now turn to the proof of Proposition \ref{prop:bialgalt}, which stated that the following equation implies strong complementarity for a pair of coherent observables $(\whiteobs, \grayobs)$.
\begin{equation}\label{eq:app-bialg-alt}
\beginpgfgraphicnamed{bialgaltbis}
\InputIfFileExists{bialgaltbis.tikz}{}{\input{./figures/bialgaltbis.tikz}}
\endpgfgraphicnamed
\end{equation}

\begin{lemma}
  Equation (\ref{eq:app-bialg-alt}) implies the following, for any pair of coherent observables:
  \begin{equation}
\beginpgfgraphicnamed{bialg_funkydirs}
\InputIfFileExists{bialg_funkydirs.tikz}{}{\input{./figures/bialg_funkydirs.tikz}}
\endpgfgraphicnamed
  \end{equation}
\end{lemma}

\begin{IEEEproof}
  \ctikzfig{bialg_funkydirs_pf1}
  
  This implies:
  \ctikzfig{bialg_funkydirs_pf2}
\end{IEEEproof}

\begin{lemma}
  Equation (\ref{eq:app-bialg-alt}) implies the following, for any pair of coherent observables:
  \begin{equation}
\beginpgfgraphicnamed{bialg_dot_copy}
\InputIfFileExists{bialg_dot_copy.tikz}{}{\input{./figures/bialg_dot_copy.tikz}}
\endpgfgraphicnamed
  \end{equation}
\end{lemma}

\begin{IEEEproof}
  \ctikzfig{bialg_dot_copy_pf1}
  \ctikzfig{bialg_dot_copy_pf2}
\end{IEEEproof}

\begin{IEEEproof}[Proof of Proposition \ref{prop:bialgalt}]
  \ctikzfig{bialg_alt_form_pf1}
  \ctikzfig{bialg_alt_form_pf2}
  
  Thus any coherent pair of observable structures satisfying Equation (\ref{eq:app-bialg-alt}) is a strongly complementary pair.
\end{IEEEproof}

\subsection{Some QM background}

In this section we present the rudiments of quantum mechanics, as
required by the main text.

The state space of a quantum system is a complex Hilbert space $X$; all
the systems we consider here will have finite dimensional  state
spaces, as is typical in quantum computation.  The possible states of
the system are unit vectors in $X$.  We use the Dirac bra-ket
notation, writing vectors as $\ket{\psi}$ and their duals as
$\bra{\psi}$.  If $X$ is $n$ dimensional, we pick a standard, or
computational, basis and write its elements $\ket{0}, \ldots
\ket{n}$.  Therefore the standard basis for the qubit---the ubiquitous
2-dimensional system used in quantum computation---is simply
$\ket{0},\ket{1}$.  Another basis often seen is the $X$-basis:
\[
\ket{+} := \frac{1}{\sqrt{2}}(\ket{0}+\ket{1})
\qquad
\ket{-} := \frac{1}{\sqrt{2}}(\ket{0}-\ket{1})\;.
\]
States with the vectors which differ only by a
complex unit factor are physically indistiguishable so the
state space is quotiented by the equivalence relation 
\[
\ket{\psi} \sim \ket{\phi} \qquad \Leftrightarrow \qquad \exists
\alpha : \ket{\psi} = e^{i\alpha}\ket{\phi}\;. 
\]
Such a scalar factor is called a \emph{global phase}.

If a quantum system is a composite of two subsystems,  having
state spaces $X$ and $Y$ respectively, then the state space of the
compound system is the \emph{tensor product} $X \otimes Y$ of the
two sub-state spaces.  When writing vectors of compound systems we may
suppress the tensor product, i.e.~$\ket{00} = \ket{0}\otimes\ket{0}$.
Notice that $\dim (X \otimes Y ) = \dim
(X) \dim(Y)$.  This has profound consequences in the form of
\emph{quantum entanglement}.  Mathematically, entanglement is the fact
that there exist states in $X \otimes Y$ which do not decompose into a
pair of states from each subsystem.  For example, if we consider two
qubits, there are no states $\ket{\psi},\ket{\phi}$ such that 
\[
\ket{00} + \ket{11} = \ket{\psi}\otimes\ket{\phi}\;.
\]
This simple mathematical fact encodes many of the counter-intuitive
properties of quantum systems, including the non-local correlations
which are the main concern of this paper, and the apparent speed-up
seen in certain quantum algorithms.

The other crucial ingredient of the quantum formalism is
\emph{measurement}.  Unlike classical systems, the quantum state may
not be observed directly, and may only be accessed via its observable
properties.  Further, not every measurement has a definite result on
every state: usually the result of a measurement will be
probabilistic.  Formally, an observable on system $X$ is a
self-adjoint operator $M:X\to X$; the possible observed values of the
observation are the eigenvalues of $M$.  We will assume throughout
that all observables have non-degerate spectra; that is we have
\[
M = \sum_{i=1}^n \lambda_i \ketbra{e_i}{e_i}
\]
where all the $\lambda_i$ are distinct.  If we perform the measurement
$M$ upon some state $\ket{\psi}$ the probability of observing
$\lambda_i$ is given the absolute square of the inner product:
\[
p(\lambda_i) = |\braket{e_i}{\psi}|^2\;.
\]
Note that since we only consider `nice' observables, we will often
identify the  observable with the basis of its eigenvectors, and speak
of, for example, ``measuring in the computational basis''.

The second shocking property of quantum measurement is the phenomenon
known as the `collapse of the wave packet'.  Less mystically put,
measurement changes the state of the system.  Precisely, after
the measurement $M$ has been performed, and the value $\lambda_i$ observed,
the state of the system is now $\ket{e_i}$, the eigenvector
corresponding to the outcome.  Therefore, repeated measurement
of the same observable will produce consistent results.  For present
purposes the actual observed values of the measurement are without
importance: the only thing that matters is the label $i$ indicating
\emph{which} outcome occurred.

As a consequence of the state changing effect of measurement, two
observables are well defined at the same time only if they have the
same eigenvectors; that is, if their respective operators commute.
Such observables are called \emph{compatible}.  In this work we are
interested in observable that are as incompatible as possible: we call
these \emph{complementary observables}. Consider the Pauli $X$ and $Z$
spin observables:
\[
\sigma_X = \begin{pmatrix} 0 & 1 \\ 1 & 0 \end{pmatrix}
\qquad
\sigma_Z = \begin{pmatrix} 1 & 0 \\ 0 & -1 \end{pmatrix}\;.
\]
Their respective eigenbases are $\ket{+},\ket{-}$ and
$\ket{0},\ket{1}$.  We have the inner products 
\begin{align*}
  |\braket{+}{0}|^2 & = |\braket{-}{0}|^2 = \frac{1}{2}\\
  |\braket{+}{1}|^2 & = |\braket{-}{1}|^2 = \frac{1}{2}
\end{align*}
Hence when one observable is well defined, the outcomes of the other
are equiprobable.  Notice that if our system is in an eigenstate of
$Z$, say, and we then measure $X$, the next measurement of $Z$ will be
completely random.  Effectively we have \emph{erased} the previous value of
of one observable by measuring the  it with a complimentary one.

The combination of entanglement and quantum measurements gives rise to
non-local correlation, which are the main subject of this paper.  As a
simple example, consider the 2-qubit entangled state known as the
\emph{Bell state},
\[
\ket{\Phi_+} = \frac{1}{\sqrt2}(\ket{00} + \ket{11}\;.
\]
Suppose that the two qubits of this state are in the hands of two
separate parties, Aleks and Bob.  If Aleks measures his qubit in the
$Z$ basis, he will observe $\ket{0}$ or $\ket{1}$ with equal
probability; suppose that it was $\ket{0}$.  The effect of Aleks's
measurement is to act on the  joint system with the operator
$\ketbra{0}{0}\otimes 1$:

\begin{multline*}
  (\ketbra{0}{0}\otimes 1)(\ket{00} + \ket{11}) =
  (\ketbra{0}{0}\otimes 1)\ket{00} + (\bra{0}\otimes 1)\ket{11} \\
  = (\ketbra{0}{0}\ket{0}\otimes \ket{0}) + (\ketbra{0}{0}\ket{1}\otimes
  \ket{1})  
  = \ket{00}\;.
\end{multline*}
(We neglect normalisation here.)  Hence, if Bob subsequently measures
his qubit he is guaranteed to observe the $\ket{0}$ outcome.
Obviously if Aleks had observed $\ket{1}$ then Bob would too; the two
parties measurements will be perfectly correlated.

Quantum systems which are not disturbed by measurement evolve
according to \emph{unitary evolution}.  That is, there is some unitary
operator $U:X\to X$ such that $\ket{\psi(t_1)} = U\ket{\psi{t_0}}$.
In quantum computation these operators are usually taken to be
composed of discrete \emph{quantum logic gates} and it is assumed that
state remains constant except when acted upon by some desired unitary
gate.  Composing such logic gates in sequence and parallel gives rise
to the \emph{quantum circuit model}.  We refer the reader to
\cite{NieChu} for details.

%\subsection{The Mermin argument}\label{App:Mermin}
%
%\bR ... schematic analysis in particular emphesising the need for determinism of the `last measurement' ... \e

\subsection{More Physical Models}
\label{sec:more-models}

While the main text takes ${\bf FHilb}$ as its primary model, there
are a variety of other categories which support some or all of the
axiomatic structures described.

For example, another $\dagger$-compact category is the category $n{\bf
  Cob}$, which has closed $(n-1)$-dimensional manifolds as objects, and
diffeomorphism classes of $n$-dimensional manifolds connecting the
objects as morphisms   \cite{Baez}.

% \footnote{\bR Q: Are there other
%   interesting models worth considering? E.g.~are there interesting
%   models that arise from constructions on categories e.g.~products and
%   other limits in Cat, spans etc.\e} 

More generally, from a given category we can construct a new model as
a subcategory.  This is a surprisingly rich source of models with
different properties.  While all the categories discussed so far have
some quantum flavour this is not necessary: modesls of classical,
reversible, and stochastic processes also arise as subcategories of
those already mentioned.

For example, the subcategory $({\bf FSet},\times)$ of $({\bf
  FRel},\times)$---defined by restricting the morphisms to
functions---can be viewed as a model of classical rather than quantum
physics.  Note that ${\bf FSet}$ is not dagger compact.  Restricting
the morphisms to permutations produces ${\bf Perm}$, the category of
reversible classical processes.  The category of stochastic classical
processes, ${\bf Stoch}$, is also a sub-theory, this time of $CP({\bf
  FHilb})$, and so is ${\bf FRel}$. In fact, as demonstrated by Coecke,
Paquette and Pavlovic in \cite{CPaqPav}, an analogue to the hierarchy
\[
{\bf Perm} \hookrightarrow {\bf FSet} 
\begin{array}{ccc}
  \mbox{\rm\rotatebox[origin=c]{45}{$\hookrightarrow$}} & \hspace{-2.5mm}\raisebox{2mm}{${\bf FRel}$}\hspace{-2.5mm} & \mbox{\rm\rotatebox[origin=c]{-45}{$\hookrightarrow$}}\\
  \mbox{\rm\rotatebox[origin=c]{-45}{$\hookrightarrow$}} & \hspace{-2.5mm}\raisebox{-2mm}{${\bf Stoch}$}\hspace{-2.5mm} & \mbox{\rm\rotatebox[origin=c]{45}{$\hookrightarrow$}}
\end{array}
CP({\bf FHilb})
\]
can be obtained for any $\dagger$-SMC ${\bf C}$.  It is therefore
possible to extract several species of `classical' processes from the
`quantum' ones, and to formulate notions like measurement, classical
data processing and classical control of quantum systems at the
abstract level of $\dagger$-SMCs, as shown in
Sec.~\ref{sec:clas-quant}.

Two more quantum-like examples obtained as subcategories are the
categories ${\bf Stab}$ \cite{CES} and ${\bf Spek}$ \cite{Spek}, which
are important models in quantum computation and quantum foundations
respectively.

\begin{definition}
Let  ${\bf Stab}$  be the subcategory of ${\bf Qubit}$ closed
under  $\dagger$-SMC structure and generated by  `observable'
$\bigl(\delta:\mathbb{C}^2\to\mathbb{C}^2\otimes\mathbb{C}^2::
|i\rangle\mapsto|ii\rangle\ , \ \epsilon:\mathbb{C}^2\to\mathbb{C}::
|i\rangle\mapsto 1\bigr)$ and  Clifford group unitaries---that is
those unitaries which preserve the eigenstates of the Pauli spin
observables---of type 
$\mathbb{C}^2\to\mathbb{C}^2$. 
\end{definition}

The category ${\bf Stab}$ represents the so-called stabilizer
restriction of quantum theory \cite{Gottesman}, which, roughly
speaking is quantum theory were we only consider six states of a
qubit, namely the X, Y and Z eigenstates.

\begin{definition}
The category   ${\bf Spek}$  is the subcategory of ${\bf FRel}_4$
closed under  $\dagger$-SMC structure and generated by `observable'
$\bigl(\delta:4\to4\times4:: 1\sim \{(1,1), (2,2)\}; 1\sim \{(1,2),
(2,1)\}; 3\sim \{(3,4), (4,3)\}; 4\sim \{(3,3), (3,3)\}\ , \
\epsilon:4\to1::\{1, 3\}\sim 1\bigr)$ and  the symmetric group
$S(4)$. 
\end{definition}

${\bf Spek}$ is a categorical presentation of the toy theory
introduced by Spekkens \cite{Spekkens} which tries to mimic ${\bf
  Stab}$ despite being \em local\em, and indeed has many features that
are usually interpreted as purely quantum. While the framework of
generalised probabilistic theories failed to capture ${\bf Spek}$ in
any significant way, both ${\bf Stab}$ and ${\bf
  Spek}$ can be easily accommodated within the process theory
framework.

Both categories are  generated by one observable (see
Sec.~\ref{sec:observables}) and a copy of $S(4)$ of unitary operations
on the smallest non-trivial system.  This in itself already
establishes that these theories bear a very close relationship, and
moreover allowed Coecke, Edwards and Spekkens to pinpoint there essential
difference \cite{CES} in terms of their \em phase groups\em, which we
define below in Sec.~\ref{sec:phasegroup}.    
%, except for the fact that it fails to be \em non-local\em, that is,
%\bR ... define non-local ...\e. 

In the case of ${\bf Stab}$ and ${\bf Spek}$ the phase group captures
their essential difference, respectively being $\mathbb{Z}_4$ and
$\mathbb{Z}_2\times\mathbb{Z}_2$ \cite{CES}.  This distinction
captures all the essential differences of the two theories, in
particular with respect to non-locality.

\end{document}

%% file: tikzfigures.tex
\newcommand{\dotcounit}[1]{%
\,\begin{tikzpicture}[dotpic,yshift=-1mm]
\node [#1] (a) at (0,0.25) {}; 
\draw [medium diredge] (0,-0.2)--(a);
\end{tikzpicture}\,}
\newcommand{\dotunit}[1]{%
\,\begin{tikzpicture}[dotpic,yshift=1.5mm]
\node [#1] (a) at (0,-0.25) {}; 
\draw [medium diredge] (a)--(0,0.2);
\end{tikzpicture}\,}
\newcommand{\dotcomult}[1]{%
\,\begin{tikzpicture}[dotpic,yshift=0.5mm]
	\node [#1] (a) {};
	\draw [medium diredge] (-90:0.55)--(a);
	\draw [medium diredge] (a) -- (45:0.6);
	\draw [medium diredge] (a) -- (135:0.6);
\end{tikzpicture}\,}
\newcommand{\dotmult}[1]{%
\,\begin{tikzpicture}[dotpic]
	\node [#1] (a) {};
	\draw [medium diredge] (a) -- (90:0.55);
	\draw [medium diredge] (a) (-45:0.6) -- (a);
	\draw [medium diredge] (a) (-135:0.6) -- (a);
\end{tikzpicture}\,}

\newcommand{\dotonly}[1]{%
\,\begin{tikzpicture}[dotpic,yshift=-0.3mm]
\node [#1] (a) at (0,0) {};
\end{tikzpicture}\,}
\newcommand{\smalldotonly}[1]{%
\,\begin{tikzpicture}[dotpic,yshift=-0.15mm]
\node [#1] (a) at (0,0) {};
\end{tikzpicture}\,}
\newcommand{\smallblackdot}{\smalldotonly{smalldot}}%NEW
\newcommand{\unit}{\dotunit{dot}}

\newcommand{\blackobs}{\ensuremath{\mathcal O_{\!\smallblackdot}}\xspace}

\newcommand{\whitedot}{\dotonly{white dot}}
\newcommand{\smallwhitedot}{\smalldotonly{smallwhitedot}}%NEW
\newcommand{\whiteunit}{\dotunit{white dot}}
\newcommand{\whitecounit}{\dotcounit{white dot}}
\newcommand{\whitemult}{\dotmult{white dot}}
\newcommand{\whitecomult}{\dotcomult{white dot}}

\newcommand{\whiteobs}{\ensuremath{\mathcal O_{\!\smallwhitedot}}\xspace}

\newcommand{\altwhitedot}{\dotonly{alt white dot}}

\newcommand{\graydot}{\dotonly{gray dot}}
\newcommand{\smallgraydot}{\smalldotonly{smallgraydot}}%NEW
\newcommand{\grayunit}{\dotunit{gray dot}}

\newcommand{\graycomult}{\dotcomult{gray dot}}

\newcommand{\grayobs}{\ensuremath{\mathcal O_{\!\smallgraydot}}\xspace}

\newcommand{\whitetranspose}{\ensuremath{{\!\!\altwhitedot\!\!}}}

\newcommand{\whiteconjugate}{\ensuremath{{\!\!\altwhitedot\!\!}}}

%% file: tikzstyles.tex
\tikzstyle{env}=[copoint,regular polygon rotate=0,minimum width=0.2cm, fill=black]

\tikzstyle{every picture}=[baseline=-0.25em]
\tikzstyle{dotpic}=[scale=0.5]
\tikzstyle{diredges}=[every to/.style={diredge}]
\tikzstyle{dot graph}=[shorten <=-0.1mm,shorten >=-0.1mm,scale=0.6]
\tikzstyle{plot point}=[circle,fill=black,minimum width=2mm,inner sep=0]

\tikzstyle{braceedge}=[decorate,decoration={brace,amplitude=2mm,raise=-1mm}]
\tikzstyle{small braceedge}=[decorate,decoration={brace,amplitude=1mm,raise=-1mm}]
\tikzstyle{left hook arrow}=[left hook-latex]
\tikzstyle{right hook arrow}=[right hook-latex]

\tikzstyle{dot}=[inner sep=0.7mm,minimum width=0pt,minimum height=0pt,fill=black,draw=black,shape=circle]
\tikzstyle{smalldot}=[inner sep=0.4mm,minimum width=0pt,minimum height=0pt,fill=black,draw=black,shape=circle]%NEW
\tikzstyle{white dot}=[dot,fill=white]
\tikzstyle{antipode}=[white dot,inner sep=0.3mm,font=\footnotesize]
\tikzstyle{smallwhitedot}=[smalldot,fill=white]%NEW
\tikzstyle{alt white dot}=[white dot,label={[xshift=3.07mm,yshift=-0.05mm,font=\footnotesize]left:$*$}]
\tikzstyle{gray dot}=[dot,fill=gray!40!white]
\tikzstyle{smallgraydot}=[smalldot,fill=gray!40!white]%NEW
\tikzstyle{box vertex}=[draw=black,rectangle]
\tikzstyle{small box}=[box vertex,fill=white]%% added rwd]
\tikzstyle{whitebg}=[fill=white,inner sep=2pt]
\tikzstyle{graph state vertex}=[sg vertex,fill=black]

\tikzstyle{wide point}=[fill=white,draw=black,shape=isosceles triangle,shape border rotate=90,isosceles triangle stretches=true,inner sep=1pt,minimum width=1.5cm,minimum height=5mm]
\tikzstyle{wide copoint}=[fill=white,draw=black,shape=isosceles triangle,shape border rotate=-90,isosceles triangle stretches=true,inner sep=1pt,minimum width=1.5cm,minimum height=4mm]
\tikzstyle{very wide copoint}=[fill=white,draw=black,shape=isosceles triangle,shape border rotate=-90,isosceles triangle stretches=true,inner sep=1pt,minimum width=2.5cm,minimum height=4mm]
\tikzstyle{very wide empty copoint}=[draw=black,shape=isosceles triangle,shape border rotate=-90,isosceles triangle stretches=true,inner sep=1pt,minimum width=2.5cm,minimum height=4mm]
\tikzstyle{symm}=[ultra thick,shorten <=-1mm,shorten >=-1mm]

\tikzstyle{square box}=[rectangle,fill=white,draw=black,minimum height=5mm,minimum width=5mm,font=\small]
\tikzstyle{square gray box}=[rectangle,fill=gray!30,draw=black,minimum height=6mm,minimum width=6mm]
\tikzstyle{point}=[regular polygon,regular polygon sides=3,draw=black,scale=0.75,inner sep=-0.5pt,minimum width=7mm,fill=white]
\tikzstyle{copoint}=[point,regular polygon rotate=180,fill=white]
\tikzstyle{gray point}=[point,fill=gray!40!white]
\tikzstyle{gray copoint}=[copoint,fill=gray!40!white]

\newcommand{\edgearrow}{{\arrow[black]{>}}}
\newcommand{\edgetick}{{\arrow[black,scale=0.7,very thick]{|}}}
\tikzstyle{diredge}=[->]
\tikzstyle{rdiredge}=[<-]
\tikzstyle{medium diredge}=[->]

\tikzstyle{short diredge}=[->]
\tikzstyle{halfedge}=[-)]
\tikzstyle{other halfedge}=[(-]
\tikzstyle{freeedge}=[(-)]
\tikzstyle{white edge}=[line width=5pt,white]
\tikzstyle{tick}=[postaction=decorate,decoration={markings, mark=at position 0.5 with \edgetick}]
\tikzstyle{small map edge}=[|-latex, gray!60!blue, shorten <=0.9mm, shorten >=0.5mm]
\tikzstyle{thick dashed edge}=[very thick,dashed,gray!40]
\tikzstyle{map edge}=[|-latex,very thick, gray!40, shorten <=1mm, shorten >=0.5mm]
\tikzstyle{tickedge}=[postaction=decorate,
  decoration={markings, mark=at position 0.5 with \edgetick}]
\tikzstyle{dirtickedge}=[postaction=decorate,
  decoration={markings, mark=at position 0.5 with \edgetick},
  decoration={markings, mark=at position 0.85 with \edgearrow}]
\tikzstyle{dirdoubletickedge}=[postaction=decorate,
  decoration={markings, mark=at position 0.4 with \edgetick},
  decoration={markings, mark=at position 0.6 with \edgetick},
  decoration={markings, mark=at position 0.85 with \edgearrow}]

\tikzstyle{probs}=[shape=semicircle,fill=gray!40!white,draw=black,shape border rotate=180,minimum width=1.2cm]

\tikzstyle{arrs}=[-latex,font=\small,auto]
\tikzstyle{arrow plain}=[arrs]
\tikzstyle{arrow dashed}=[dashed,arrs]
\tikzstyle{arrow bold}=[very thick,arrs]
\tikzstyle{arrow hide}=[draw=white!0,-]
\tikzstyle{arrow reverse}=[latex-]
\tikzstyle{cdnode}=[]

\tikzstyle{cnot}=[fill=white,shape=circle,inner sep=-1.4pt]
\tikzstyle{wire label}=[font=\tiny, auto]

%% file: defs.tex
\def\whitemu{\mu_{\smallwhitedot}}

\def\whiteeta{\eta_{\smallwhitedot}}

\def\graydelta {\delta_{\smallgraydot}}
\def\whitedelta{\delta_{\smallwhitedot}}

\def\grayepsilon {\epsilon_{\smallgraydot}}
\def\whiteepsilon{\epsilon_{\smallwhitedot}}

\def\whitePhi{\Phi_{\!\smallwhitedot}}
\def\graymeas{m_{\!\smallgraydot}}

\newcommand{\bra}[1]{\ensuremath{\left\langle #1 \right|}}
\newcommand{\ket}[1]{\ensuremath{\left|  #1 \right\rangle}}
\newcommand{\roundket}[1]{\ensuremath{\left|  #1 \right)}}
\newcommand{\braket}[2]{\ensuremath{\langle#1|#2\rangle}}
\newcommand{\ketbra}[2]{\ensuremath{\ket{#1}\!\bra{#2}}}

\newcommand{\catC}{\ensuremath{\mathcal{C}}\xspace}
\newcommand{\catCop}{\ensuremath{\mathcal{C}^{\mathrm{op}}}\xspace}

\newcommand{\catFHilb}{\ensuremath{\textrm{\bf FHilb}}\xspace}

\tikzstyle{cdiag}=[matrix of math nodes, row sep=3em, column sep=3em, text height=1.5ex, text depth=0.25ex,inner sep=0.5em]
\tikzstyle{arrow above}=[transform canvas={yshift=0.5ex}]
\tikzstyle{arrow below}=[transform canvas={yshift=-0.5ex}]

%% file: scalar.tikz
\begin{tikzpicture}
	\begin{pgfonlayer}{nodelayer}
		\node [style=gray dot] (0) at (0, 0.12) {};
		\node [style=white dot] (1) at (0, -0.12) {};
	\end{pgfonlayer}
	\begin{pgfonlayer}{edgelayer}
		\draw [style=none] (1) to (0);
	\end{pgfonlayer}
\end{tikzpicture}

%% file: classicalpointaction.tikz
\begin{tikzpicture}
	\begin{pgfonlayer}{nodelayer}
		\node [style=none] (0) at (-1, 0.15) {};
		\node [style=white dot] (1) at (-1, -0.12) {\footnotesize $\!\!i\!\!$};
		\node [style=none] (2) at (-1, -0.4) {};
	\end{pgfonlayer}
	\begin{pgfonlayer}{edgelayer}
		\draw (1) to (2.center);
		\draw (0.center) to (1);
	\end{pgfonlayer}
\end{tikzpicture}

%% file: classicalpointwhite.tikz
\begin{tikzpicture}
	\begin{pgfonlayer}{nodelayer}
		\node [style=none] (0) at (0, 0.25) {};
		\node [style=white dot] (1) at (0, 0) {\footnotesize $\!\!i\!\!$};
	\end{pgfonlayer}
	\begin{pgfonlayer}{edgelayer}
		\draw (0.center) to (1);
	\end{pgfonlayer}
\end{tikzpicture}

%% file: innerprod.tikz
\begin{tikzpicture}[dotpic]
	\begin{pgfonlayer}{nodelayer}
		\node [style=dot] (0) at (-4, 0.25) {};
		\node [style=gray dot] (1) at (-4, -0.25) {};
	\end{pgfonlayer}
	\begin{pgfonlayer}{edgelayer}
		\draw [style=short diredge] (1) to (0);
	\end{pgfonlayer}
\end{tikzpicture}

%% file: bialg-scalar.tikz
\begin{tikzpicture}
	\begin{pgfonlayer}{nodelayer}
		\node [style=white dot] (0) at (-0.5, 0.25) {};
		\node [style=none] (1) at (0, 0) {$=$};
		\node [style=white dot] (2) at (-0.5, -0.25) {};
	\end{pgfonlayer}
	\begin{pgfonlayer}{edgelayer}
		\draw [style=diredge] (2) to (0);
	\end{pgfonlayer}
\end{tikzpicture}